\documentclass[pre]{revtex4-1}
\pdfoutput=1
\usepackage{amsmath,amssymb,latexsym,epsfig,graphics,epsf}
\usepackage{graphics}
\usepackage{float}

\newcommand{\Ginf}{G_\infty}
\newcommand{\be}{\begin{equation}}
\newcommand{\ee}{\end{equation}}
\newcommand{\fig}[1]{Fig.~\ref{#1}}
\newcommand{\eq}[1]{Eq.~(\ref{#1})}
\renewcommand{\sec}[1]{Sec.~\ref{#1}}

\begin{document}
\title{Model for the alpha and beta shear-mechanical properties of supercooled liquids and its comparison to squalane data}
\author{Tina Hecksher, Niels Boye Olsen}
\author{Jeppe C. Dyre}\email{dyre@ruc.dk}
\affiliation{``Glass and Time'', IMFUFA, Dept. of Science and Environment, Roskilde University, P. O. Box 260, DK-4000 Roskilde, Denmark}

\date{\today}

\begin{abstract}
This paper presents data for supercooled squalane's frequency-dependent shear modulus covering frequencies from 10 mHz to 30 kHz and temperatures from 168 K to 190 K; measurements are also reported for the glass phase down to 146 K. The data reveal a strong mechanical beta process, also above the glass transition. A model is proposed for the shear response of the metastable equilibrium liquid phase of supercooled liquids. The model is an electrical equivalent-circuit characterized by additivity of the dynamic shear compliances of the alpha and beta processes. The nontrivial parts of the alpha and beta processes are represented by a ``Cole-Cole retardation element'' defined as a series connection of a capacitor and a constant-phase element, resulting in the Cole-Cole compliance function well-known from dielectrics. The model, which assumes that the high-frequency decay of the alpha shear compliance loss varies with angular frequency as $\omega^{-1/2}$, has seven parameters. Assuming time-temperature superposition for the alpha and the beta processes separately, the number of parameters varying with temperature is reduced to four. The model provides a better fit to data than a seven-parameter Havriliak-Negami type model. From the temperature dependence of the best-fit model parameters the following conclusions are drawn: 1) the alpha relaxation time conforms to the shoving model; 2) the beta relaxation loss-peak frequency is almost temperature independent; 3) the alpha compliance magnitude, which in the model equals the inverse of the instantaneous shear modulus, is only weakly temperature dependent; 4) the beta compliance magnitude decreases by a factor of three upon cooling in the temperature range studied. The final part of the paper briefly presents measurements of the dynamic adiabatic bulk modulus covering frequencies from 10 mHz to 10 kHz in the temperature range 172 K to 200 K. The data are qualitatively similar to the shear data by having a significant beta process. A single-order-parameter framework is suggested to rationalize these similarities.
\end{abstract}

\maketitle

\section{Introduction}

Many organic liquids are easily supercooled and excellent glass formers, usually with the glass transition taking place far below room temperature. Such systems are experimentally convenient for studying the physics of highly viscous liquids, the glass transition, glassy relaxation, etc, phenomena that are believed to be universal for basically all liquids \cite{har76,bra85,edi96,ang95,dyr06,ber11,flo11,sti13}. As the liquid is cooled, the relaxation time and viscosity increase by many orders of magnitude over a narrow temperature range. Beyond the dominant and slowest alpha relaxation process many liquids have additional faster relaxation(s), notably the so-called beta relaxation. The alpha and beta processes are often studied by means of dielectric spectroscopy. They are also present, however, in the liquid's mechanical properties, which are the focus of the present paper presenting squalane data and a model for supercooled liquids' dynamic shear-mechanical properties.

Squalane is a liquid alkane consisting of a linear ${\rm C}_{24}$ backbone with six symmetrically placed methyl groups. Its systematic name is 2,6,10,15,19,23-hexamethyltetracosane. Squalane is a van der Waals liquid that is an excellent glass former \cite{har76,dee99,ric03,jak05,com14}. Squalane's melting point is $T_m=235$ K and its glass transition temperature $T_g\cong 168$K follows the well-known rule $T_g\sim (2/3)T_m$ \cite{ang00,dyr06}. Squalane has low toxicity and is used in cosmetics as moisturizer; due to the complete saturation squalane is not subject to auto-oxidation \cite{wiki}.  In basic research squalane is used as reference liquid in tribology and for elucidating the mechanism of elastohydrodynamic friction \cite{bai11,com13,spi14,sch15a}. Squalane has been studied in molecular dynamics simulations of nonlinear flows \cite{bai02}. Squalane has also been used as a solvent for studying the intriguing Debye dielectric relaxation of mono-hydroxy alcohols \cite{wan05}, the rotation of aromatic hydrocarbons in viscous alkanes \cite{bro99}, and the Stokes-Einstein relation for diffusion of organic solutes \cite{kow11}. Due to its low vapor pressure squalane is used as a benchmark molecule for reaction-dynamics experiments performed under ultrahigh vacuum \cite{sae91,koh06}.

Measurements of neat supercooled squalane's dynamic shear modulus in the MHz range were reported many years ago \cite{bar72}. Subsequent studies of squalane include measurements of its dielectric relaxation \cite{ric03} and dynamic shear modulus over frequencies ranging from a few mHz to 10 Hz \cite{dee99}, later extended to 30 kHz \cite{jak05}. The present paper covers the latter range of frequencies with more accurate data and for more temperatures than Ref. \onlinecite{jak05}. The main motivation is not to present new data, however, but to introduce an electrical equivalent-circuit model representing data very well. The model is a modification of one discussed previously by our group, which introduces a crucial extra capacitor \cite{jak11}. 

Section \ref{data_sec} presents the squalane data and the piezo-ceramic transducer used to obtain them. Section \ref{equiv_cir} introduces electrical-equivalent circuit modeling of linear mechanical relaxation phenomena in general and motivates the model. It has four free parameters of dimensions and three dimensionless ``shape'' parameters that are fixed from fitting to data at one temperature. Section \ref{fit_sec} shows that the model fits data very well, considerably better than a similar Havriliak-Negami type model with the same number of parameters. While the paper's main focus is on the dynamic shear data, Sec. \ref{bulk_sec} supplements these by presenting dynamic adiabatic bulk-modulus data. It is briefly shown that these may interpreted in terms of an electrical equivalent circuit model in which the dissipation is controlled by the dynamic shear modulus. Finally, Sec. \ref{disc_sec} gives a discussion with a focus on the temperature dependence of the best-fit model parameters, showing that these conform to the shoving model and that the beta process activation energy is temperature independent. If these two findings were built into the model, it would have just two parameters varying with temperature.

\section{Data for the dynamic shear modulus of squalane}\label{data_sec}

\begin{figure}
  \includegraphics[height=6cm]{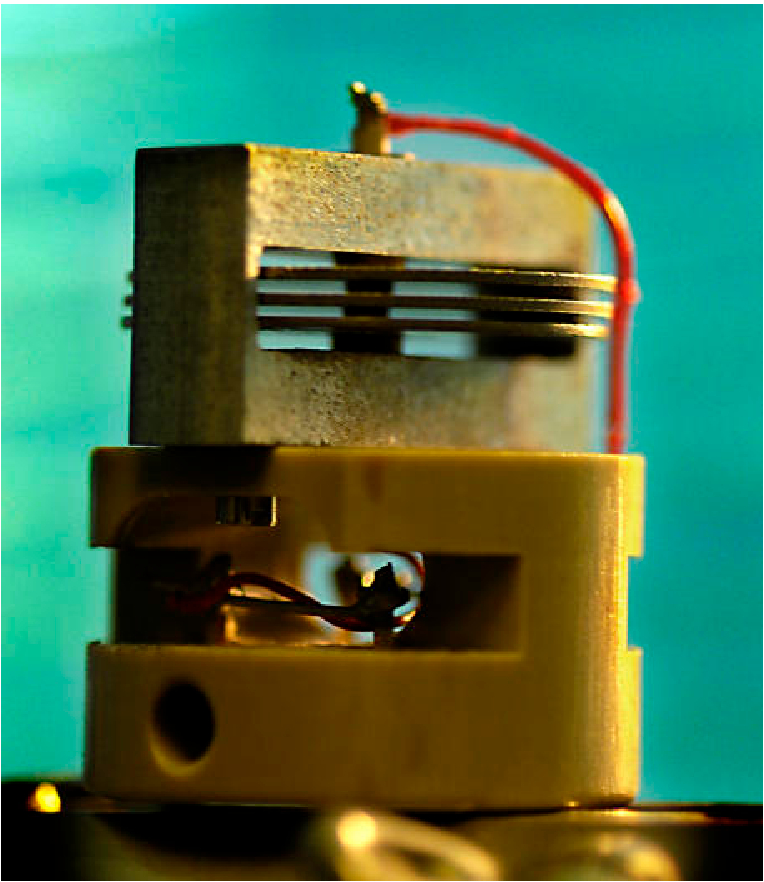}
    \includegraphics[height=6cm]{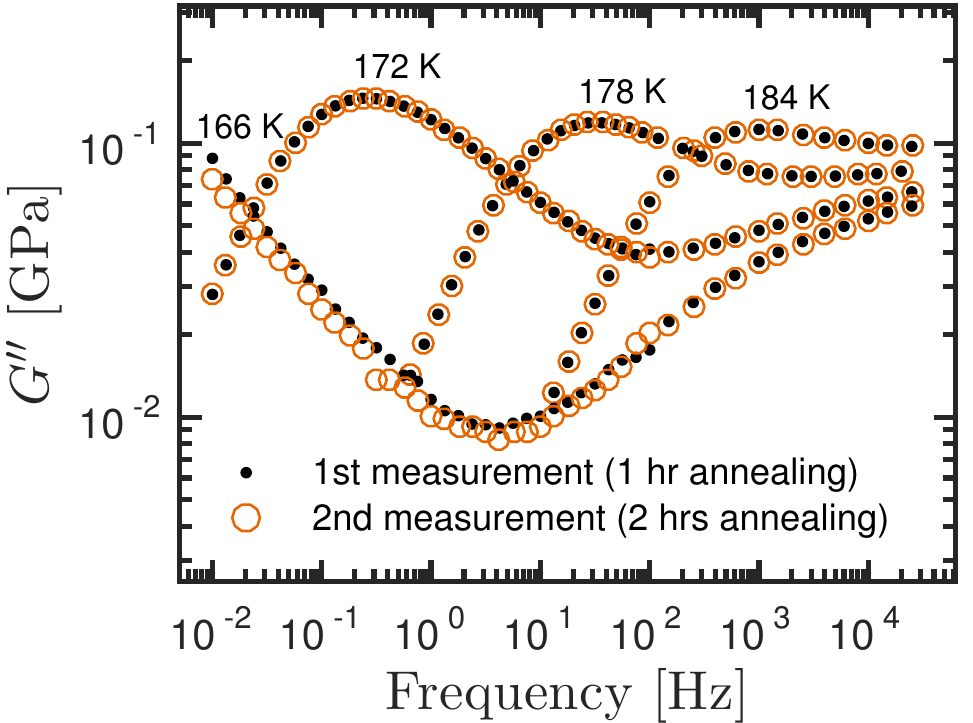}
  \caption{(a) Three-disc transducer used for measuring the dynamic shear modulus over frequencies from 1 mHz to 50 kHz above which resonances make it impossible to measure in the quasi-static mode \cite{chr95}; this paper presents data from 10 mHz to 30 kHz. The discs are polarized piezo-ceramics with electrodes on both sides that are electrically coupled in such a way that when the middle disc expands radially, the upper and lower discs contract by half the amount \cite{chr95}. Each disc has thickness 0.5 mm and diameter 20 mm. The liquid is placed between discs 1 and 2 and between discs 2 and 3 at room temperature at which squalane has low viscosity. 
  (b) Examples of data for the imaginary part of the shear modulus as a function of frequency, $G''(\omega)$, illustrating the measurement protocol: Starting at a high temperature data were generated by moving in steps of 2 K down to 146 K. At each temperature the sample was equilibrated for one hour before measuring, which  takes approximately another hour (in the present case when the lowest frequency is 10 mHz). The procedure was repeated to ensure reproducibility and that the sample is in (metastable) thermal equilibrium. The 166 K data show that equilibrium was not reached. We do not analyze data below 168 K, but above this temperature the liquid is in a state of metastable equilibrium and there is full reproducibility.\label{PSG}
   }
\end{figure}

This paper focuses on the modeling of the dynamic shear-mechanical properties of metastable equilibrium supercooled liquids, \textit{in casu} squalane above its glass transition temperature 168 K. Measurements were performed at temperatures down to 146 K, however, which is well into the glass. Data were obtained with 2 K intervals using the three-disk piezo-ceramic shear transducer shown in \fig{PSG}(a) \cite{chr95} in the setup described in Ref. \onlinecite{hec13}. The cryostat keeps temperature stable within 10 mK. References \onlinecite{iga08a} and \onlinecite{iga08b} give details about the home-built cryostat and impedance-measuring setup.

Before measuring, the filled transducer was annealed at the highest temperature for 30 hours in order to equilibrate the ceramics. After this, with 2 K intervals several temperatures were monitored by first equilibrating for one hour, after which a frequency spectrum was measured which  lasted approximately one hour. This measurement was repeated to ensure reproducibility, i.e., that the liquid is in metastable equilibrium and that the setup works properly. All in all, approximately three hours were spent at each temperature. The protocol is illustrated in \fig{PSG}(b). After all measurements had finished, the empty transducer was calibrated \cite{hec13}. If everything works, a set of data as those analyzed below may be obtained within less than a week.

\begin{figure}[h!]  
  \includegraphics[width=\textwidth]{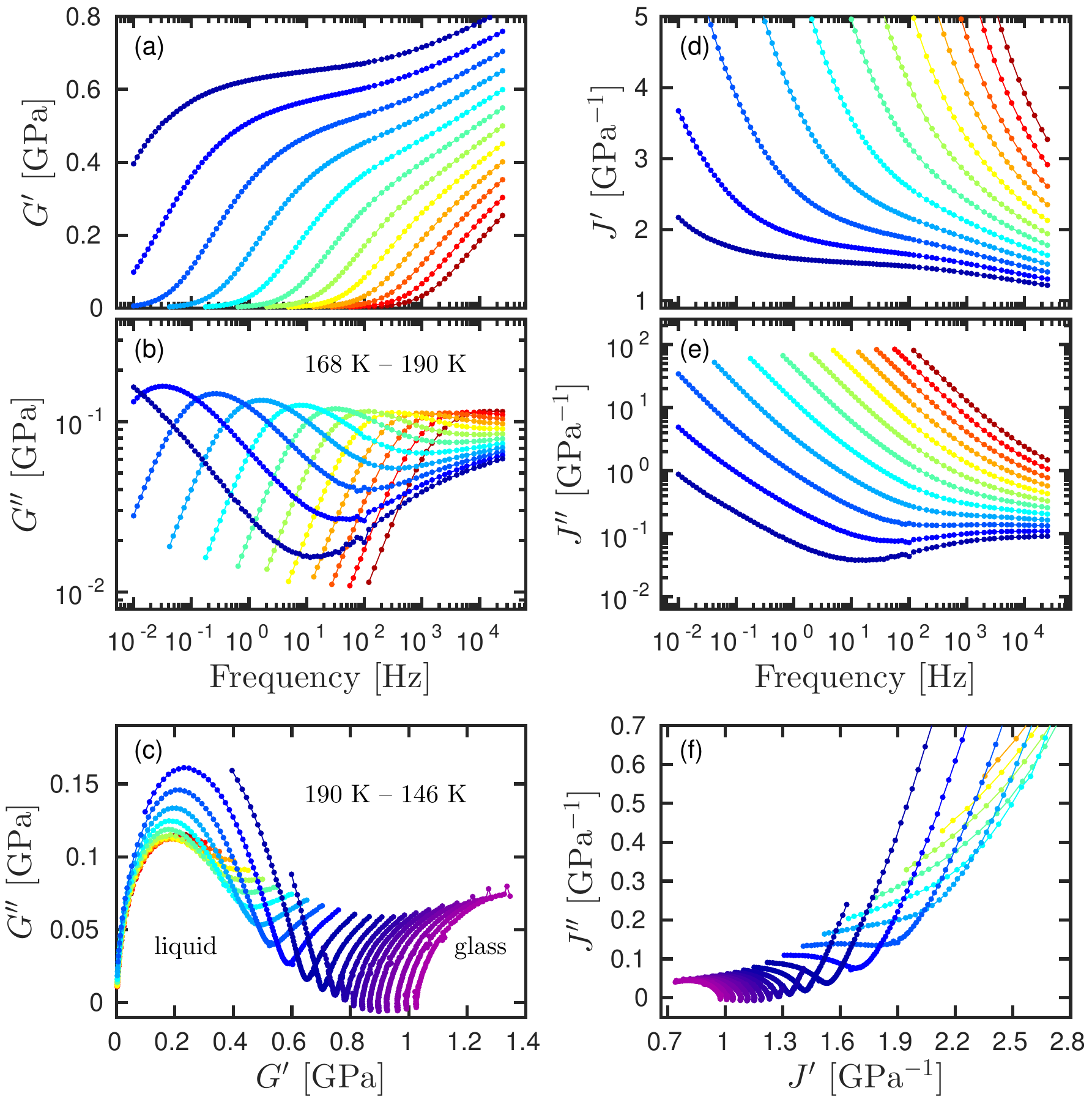}
  \caption{The dynamic shear modulus $G(\omega)$ and shear compliance $J(\omega)=1/G(\omega)$ of the metastable equilibrium supercooled liquid phase of squalane for temperatures ranging from 168 K to 190 K probed at intervals of 2 K \cite{datarepository}. The blue symbols correspond to low temperatures, the red ones to high temperatures.
  	(a) and (b) show the real and imaginary parts of $G(\omega)\equiv G'(\omega)+iG''(\omega)$.
  	(c) shows a so-called Nyquist plot of $G(\omega)$ plotting the real part along the $x$ axis and the imaginary part along the $y$ axis in a curve parametrized by the frequency; we here included data extending into the glassy state where only the $\beta$ process is observable.
    (d) and (e) show the real and negative imaginary parts of $J(\omega)\equiv J'(\omega)-iJ''(\omega)$; 
    (f) shows the Nyquist plot of $J(\omega)$ (excluding the highest temperatures).\label{G_data} 
   }
\end{figure}

Figures \ref{G_data}(a) and (b) present the real and imaginary parts of squalane's dynamic shear modulus $G(\omega)$ in which $\omega$ is the angular frequency. Shown in \fig{G_data}(c) is a so-called Nyquist plot of  $G(\omega)$, i.e., the real versus imaginary parts parametrized by the frequency; in this figure data for the glassy phase were included. A strong beta relaxation is observed. Figures \ref{G_data}(d) and (e) present the real and imaginary parts of the dynamic shear compliance $J(\omega)=1/G(\omega)$, while \fig{G_data}(f) gives the Nyquist plot of $J(\omega)$.

\section{Electrical-equivalent circuit model}\label{equiv_cir}

\subsection{Philosophy of circuit modeling}

Some scientists regard the modeling of linear-response data by an electrical equivalent circuit as old-fashioned. A common argument is that all data may be fitted by an electrical circuit and that, consequently, such type of models can contribute little if physical insight is the goal. In our opinion this is not quite correct \cite{sag10}, and the following reasons may be given for using electrical equivalent circuits for rationalizing data as a first step towards a physical understanding:

\begin{itemize}

\item {\it Physical consistency.}  Circuit models guarantee that inconsistencies are avoided. Not only is {\it linearity} ensured, so is {\it causality} and {\it positive dissipation},  requirements that any linear-response model must obey. 

\item {\it Simplicity.} The electrical circuit defines the model. Even simple circuits represent several differential equations.

\item {\it Same language as that used for modeling the experimental setup.} This paper is concerned with the interpretation of linear-response data for the dynamic shear modulus. The experimental setup used for obtaining the data is itself modeled by an electrical equivalent circuit \cite{iga08a,iga08b,hec13}, and there is an element of economy in using electrical equivalent circuits for both purposes.

\item {\it High- and low-frequency limits.} These limits are straightforward to identify for a given circuit.

\item {\it Couplings between different linear-response functions are easily introduced.} For glass-forming liquids a major challenge is to understand the relation between different frequency-dependent linear-response functions like the shear  and bulk moduli, dielectric constant, specific heat, etc \cite{jak12}. Such relations are conveniently modeled via electrical equivalent circuits related by transformers in which, for instance, charge is transformed into mechanical displacement / entropy, electrical current into velocity (shear rate) / heat current over temperature, and voltage into mechanical force / temperature \cite{pvc78,systemdyn}. Shifts between different types of variables are represented by transformer elements with the property that the power -- the product of generalized ``voltage'' and ``current'' -- is invariant. For more on the general ``energy-bond'' formalism the reader is referred to Refs. \onlinecite{paynter,ost71,pvc78,systemdyn,III}; a brief discussion is given below.

\item {\it Straightforward extension to a nonlinear model via parametric control.} An electrical equivalent circuit's parameters vary with the thermodynamic state point. Having such dependencies controlled by charges or voltages at particular points of the circuit opens for constructing fairly simple models of physical aging, which automatically ensure that no fundamental physical laws are violated. 

\end{itemize}

Once an electrical equivalent circuit has been constructed representing data accurately, this provides an important input for constructing a microscopic physical model of the system in question. We regard the circuit as a help towards eventually obtaining the ultimate microscopic understanding, not as the final model itself.

\subsection{Basic circuit elements}

Rheology has its own circuit language based on dashpots representing Newtonian viscous flow and springs representing a purely elastic response \cite{ferry}. This language is mathematically equivalent to that of electrical circuit modeling, and which language to use is a matter of convenience. As physicists we are more used to electrical circuits. Their use has the additional advantage of easily relating to dielectric relaxation phenomena, which are of great importance for glass-forming liquids \cite{ric15} and experimentally closely connected to the shear-mechanical properties \cite{dim74,nis05,buc11,gar15a}.

Translating from electrical to rheological circuits is a bit counterintuitive when it comes to the  diagrammatic representation because series connections become parallel connections and \textit{vice versa}: two elements in series in an electrical circuit have the \textit{same current}, which corresponds to the analogous rheological elements being placed \textit{in parallel} because the two shear displacements are identical. Likewise, an electrical-circuit parallel connection translates into a mechanical series connection. Once this is kept in mind, however, translation between the two languages is straightforward and unique.

Since electrical equivalent circuits are used for modeling dynamic mechanical relaxation phenomena, we shall not distinguish between the dynamic capacitance $C(\omega)$ and the dynamic shear compliance $J(\omega)$. There is as mentioned a general circuit modeling language -- the energy-bond graph formalism \cite{paynter,ost71,pvc78,systemdyn} -- which may be used also, e.g., for thermal relaxation phenomena. A general energy-bond is characterized by an ``effort'' variable $e$ and a ``flow'' variable $f$, the product of which gives the power transferred into the system from its surroundings. For instance, for a thermodynamic energy-bond $e$ is the temperature deviation from a reference temperature and $f$ is the entropy current, i.e., the heat flow over temperature \cite{meixner,pvc78,ber64}.

How to translate electrical linear-response functions to the corresponding rheological ones? With the energy-bond formalism in mind, the displacement $q$ represents electrical charge or shear displacement (strain), the flow given by $f\equiv\dot q$ represents electrical current or shear rate, and the effort $e$ represents voltage drop or shear stress \cite{paynter,ost71,pvc78}. 

The most important complex-valued linear-response functions translate as follows when given as functions of the angular frequency $\omega$ in the standard way, e.g.,  $q(t)=\rm{Re}\left[q(\omega)\exp(i\omega t)\right]$ in which $q(\omega)$ is the complex amplitude: 

\begin{itemize}

\item Electrical capacitance $C(\omega)$ corresponds to the complex shear compliance $J(\omega)$ since both in the Fourier domain are equal to displacement/effort, i.e., $q(\omega)/e(\omega)$. If the symbol $\sim$ is used for ``corresponds to'', this is summarized follows:

\be\label{CJeq}
C(\omega)
\,\sim\,J(\omega)
\,\sim\,\frac{q(\omega)}{e(\omega)}\,.
\ee	
	
\item Inverse electrical capacitance $1/C(\omega)$ corresponds to the complex shear modulus $G(\omega)=1/J(\omega)$ since both in the Fourier domain are equal to effort/displacement, i.e., $e(\omega)/q(\omega)$:

\be
\frac{1}{C(\omega)}
\,\sim\,G(\omega)
\,\sim\,\frac{e(\omega)}{q(\omega)}\,.
\ee

\item Electrical impedance $Z(\omega)=1/Y(\omega)$ corresponds to the complex shear viscosity $\eta(\omega)$ since both in the Fourier domain are equal to effort/flow, i.e., $e(\omega)/f(\omega)=e(\omega)/\dot q(\omega)=e(\omega)/(i\omega q(\omega))$:

\be
Z(\omega)
\,\sim\,{\eta(\omega)}
\,\sim\,\frac{e(\omega)}{i\omega q(\omega)}\,.
\ee

\end{itemize}

\begin{figure}[h!]  
  \includegraphics[width=7cm]{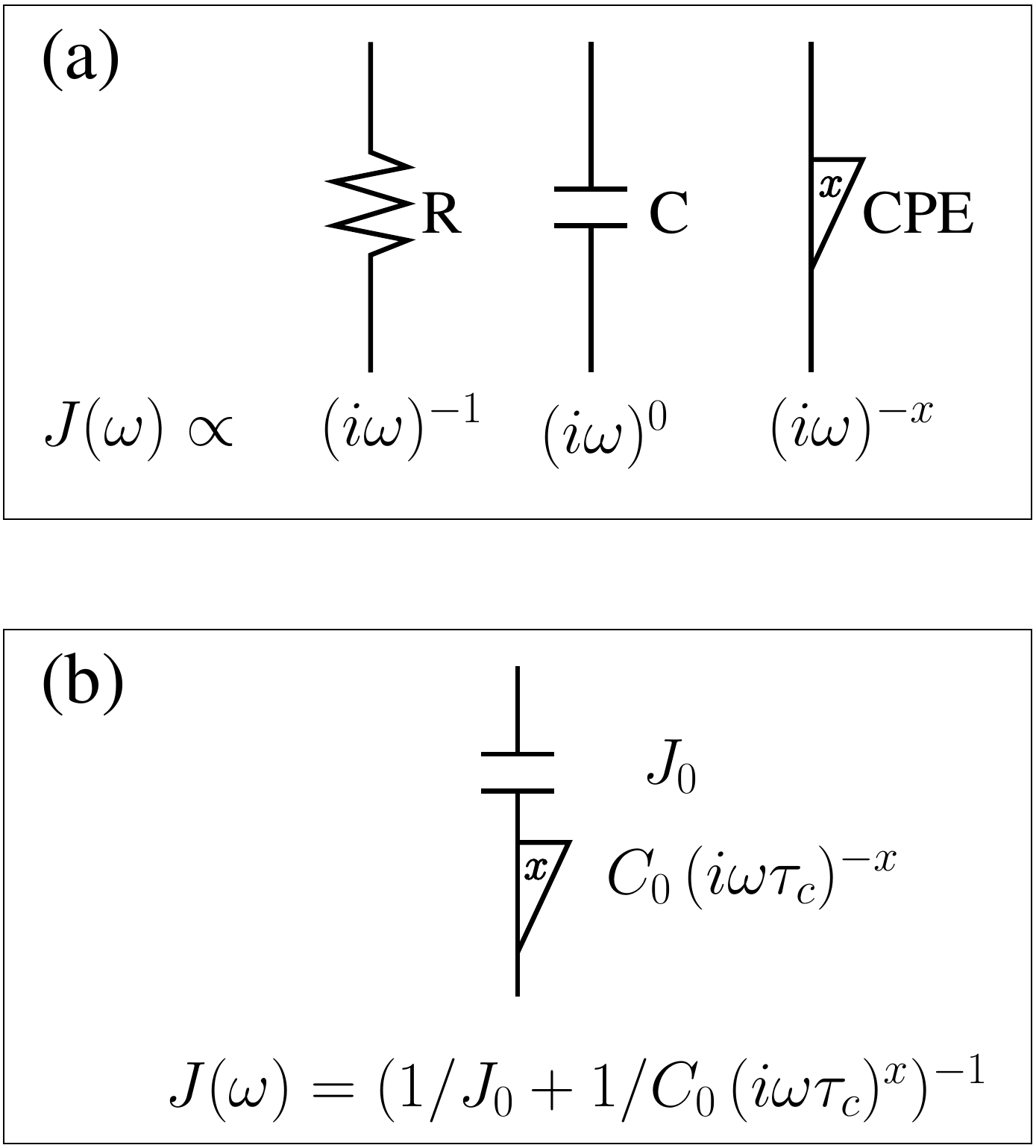}
  \caption{Electrical equivalent-circuit modeling.
  (a) The three basic elements: a resistor (R), a capacitor (C), and a constant-phase element (CPE). Their complex, frequency-dependent capacitances -- corresponding to the dynamic shear compliance $J(\omega)$ (\eq{CJeq}) -- vary with angular frequency $\omega$ in proportion to, respectively, $(i\omega)^{-1}$, $(i\omega)^{0}={\rm Const.}$, and $(i\omega)^{-x}$ ($0<x<1$).
  (b) The Cole-Cole retardation element (CCRE) leading to the well-known Cole-Cole expression for the capacitance (shear compliance)  \cite{col41}, \eq{CC}.
   \label{CIR1}}
\end{figure}

Three basic elements are used below (\fig{CIR1}(a)): resistors, capacitors, and constant-phase elements (CPE) \cite{jonscher}. A CPE is characterized by a capacitance that as a function of $\omega$ varies as 

\be\label{CPE}
C(\omega)\propto (i\omega)^{-x}
\ee
in which $0<x<1$. The name CPE reflects the fact that the ratio between the real and imaginary parts of $C(\omega)$ is the same at all frequencies, which implies a constant phase difference between displacement and effort. The CPE is a generalization of capacitors and resistors because a capacitor obeys $C(\omega)\propto (i\omega)^0$= Const. while a resistor's capacitance is given by $C(\omega)\propto (i\omega)^{-1}$. Thus allowing for $0\le x\le 1$ there is just a single ``Lego block'' in the model tool box, namely the CPE. -- Note that \eq{CPE} translates into 

\begin{itemize}
	\item $J(\omega)\propto(i\omega)^{-x}$ for the CPE dynamic shear compliance,
	\item $G(\omega)\propto(i\omega)^{x}$ for the CPE dynamic shear modulus,
	\item $\eta(\omega)\propto (i\omega)^{x-1}$ for the CPE dynamic shear viscosity.
\end{itemize}

\subsection{Parametrizing the constant-phase element}

For the CPE we define a magnitude constant $C_0$ and a characteristic time $\tau_c$ by writing

\be\label{CPE2}
C(\omega)
\,=\, C_0\, (i\omega\tau_c)^{-x}\,.
\ee
Because the CPE is time-scale invariant, the constants $C_0$ and $\tau_c$ are not uniquely determined since the same physics is described by using instead for any number $k>0$ the magnitude constant $kC_0$ and the characteristic time $k^{1/x}\tau$. The CPE is central for the model proposed below, and for the discussion of the model parameters' temperature dependence in fit to data (Sec. \ref{disc_sec}) we need a convention about the magnitude constant and the characteristic time. We take $C_0$ to be a \textit{universal, temperature-independent} number. The motivation is that, if any physics is to be ascribed to $\tau_c$, the CPE magnitude constant $C_0$ should also make sense physically. Since squalane like most other organic glass-forming liquids have an instantaneous shear modulus, i.e., high-frequency plateau shear modulus, of order GPa, we fix the magnitude constant as follows:

\be\label{gpainv}
C_0
\,\equiv\, 1\, {\rm GPa}^{-1}\,.
\ee

\subsection{The Cole-Cole retardation element}

What is here termed a Cole-Cole retardation element (CCRE) consists of a series connection of a CPE and a capacitor (\fig{CIR1}(b)). Recall that the capacitance $C(\omega)$ of two elements in series with capacitances $C_1(\omega)$ and $C_2(\omega)$ is given by $1/C(\omega)=1/C_1(\omega)+1/C_2(\omega)$. Thus if the CCRE capacitor's value is $J_0$, the CCRE compliance $J(\omega)$ is given by

\be\label{CCRE}
\frac{1}{J(\omega)}
\,=\,\frac{1}{J_0}+\frac{1}{C_0(i\omega\tau_c)^{-x}}\,.
\ee
The CCRE is named after the Cole-Cole dielectric capacitance function from 1941 \cite{col41}, which in the mechanical language is the following expression

\be\label{CC}
J(\omega)
\,=\,  \frac{J_0}{1+(i\omega\tau)^x}\,.
\ee
Here $J_0$ is the DC shear compliance and $\tau$ the inverse angular loss-peak frequency. It is straightforward to show that \eq{CCRE} leads to \eq{CC} if one identifies

\be\label{taueq}
\tau
\,\equiv\,\tau_c\,\left(\frac{J_0}{C_0}\right)^{1/x}\,.
\ee
Note that the characteristic time is \textit{not} identical to the inverse loss-peak frequency of the CCRE. 

The fit to data (\sec{fit_sec}) gives CPE characteristic times $\tau_c$ that are thermally activated for both the alpha and the beta process. As demonstrated in \fig{activE} below, the alpha CPE characteristic time activation energy is proportional to the instantaneous shear modulus, whereas the beta CPE characteristic time activation energy is virtually temperature independent. Physically, we think of each CPE's characteristic time $\tau_c$ as reflecting this element's ``inner'' clock somewhat analogous to the material time of the Narayanaswamy physical-aging theory \cite{nar71,scherer,hec15,dyr15}. 

\begin{figure}[h!]  
	\includegraphics[width=7cm]{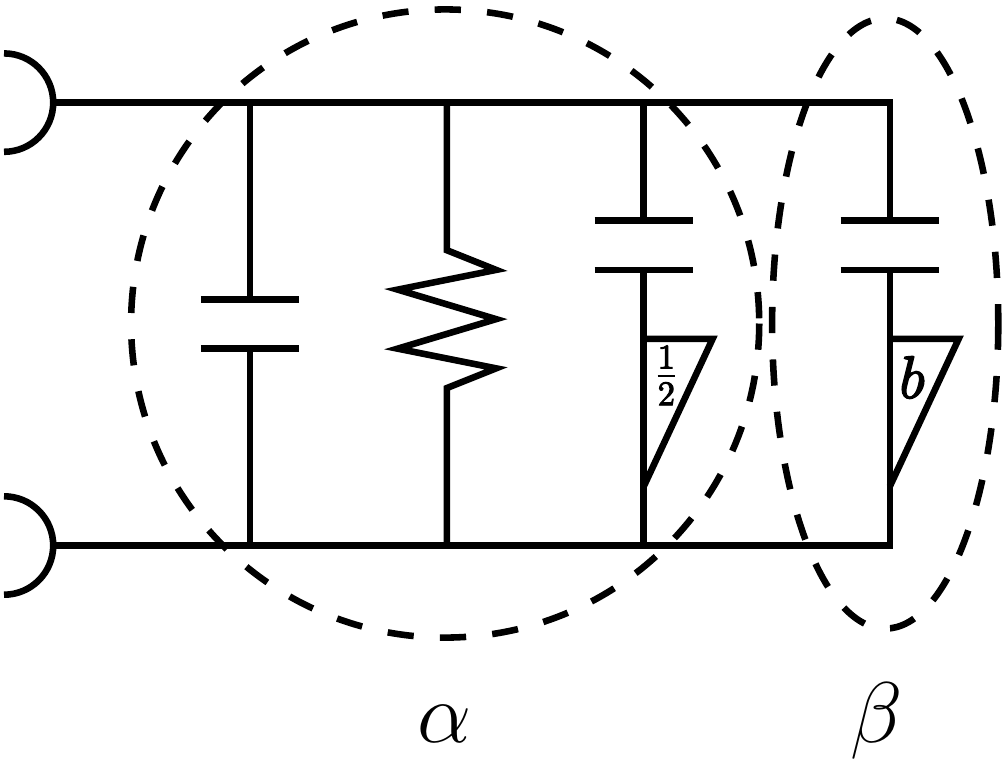}
	\caption{The electrical equivalent-circuit model for the dynamic shear-mechanical properties of a supercooled liquid. The model is characterized by additivity of the alpha and beta compliances. The alpha process is represented by a standard RC element -- corresponding to the classical Maxwell model \eq{Max} -- in parallel with a Cole-Cole retardation element (CCRE) with exponent $1/2$. The beta process is represented by an additional CCRE. The model has seven parameters, one for each basic element except for the beta CPE element that has two parameters. In the fit to squalane data three dimensionless parameters are assumed to be temperature independent, corresponding to the assumption that the alpha and beta shear compliances separately obey time-temperature superposition. 
	}\label{CIR2}
	\end{figure}

\subsection{Model for the dynamic shear-mechanical properties}

To arrive at an electrical equivalent circuit model of a supercooled liquid's shear dynamic properties, we first note that a standard parallel electrical RC element corresponds to the classical Maxwell model for viscoelasticity. This beatifully simple model is based on the assumption that the stress decays exponentially to zero whenever the sample is at rest \cite{har76,lam78,dyr06}. If the time-dependent shear stress is denoted by $\sigma(t)$, the shear displacement (strain) by $\gamma$, the DC shear viscosity by $\eta_0$, and the instantaneous shear modulus by $G_\infty$, the differential equation describing the Maxwell model for an arbitrary shear rate as a function of time, $\dot{\gamma}(t)$, is \cite{har76,lam78,dyr06}

\be\label{Max}
\dot{\gamma}(t)=\frac{\sigma(t)}{\eta_0}+ \frac{\dot{\sigma}(t)}{G_\infty}\,.
\ee
The Maxwell model is equivalent to a parallel electrical RC element because for such an element the voltage is the same across both resistance and capacitor, which in the Maxwell model corresponds to having the same shear stress. The resistor current corresponds to the first term on the right hand side of \eq{Max} (a dashpot in the traditional language of viscoelasticity), and the capacitor current corresponds to the second term (a spring in the viscoelastic language).

The Maxwell model is too simple to fit data for glass-forming liquids, however, and must be extended by including one or more non-trivial dissipative terms. This paper's basic idea is that these terms are described by CCREs placed in parallel to the Maxwell RC element, one for the alpha process and one for the beta process (\fig{CIR2}).

In the model none of the two CCREs are inherently linked to the alpha process RC element. Nevertheless, one CCRE will be regarded as part of the alpha process for the following reasons. Previous publications of the {\it Glass and Time} group have presented experimental \cite{ols01,nie09,hec13} and theoretical \cite{dyr05,dyr06a,dyr07} evidence that in the absence of beta relaxation the alpha process has a generic $\omega^{-1/2}$ high-frequency decay of the dielectric loss and shear compliance. This is an old idea that keeps resurfacing, recently in an interesting biophysical context \cite{oos16}, and a generic $\omega^{-1/2}$ high-frequency decay is the characteristic feature of the 1967 Barlow-Erginsav-Lamb (BEL) model \cite{BEL,bar72a,har76,lam78,dyr05}. In view of this we fix the exponent to $1/2$ for one CCRE and regard this element as a part of the alpha process. Confirming this assignment, for liquids without a mechanical beta relaxation like the silicone diffusion pump oils DC704 and DC705, the dynamic shear compliance is well fitted by the model of \fig{CIR2} without the beta CCRE (unpublished). It should be mentioned that a finite one-dimensional so-called diffusion chain describing, e.g., the relation between temperature and heat flux entering from one end, has a compliance function that is very close to that of the alpha part of the model in \fig{CIR2}.

The dynamic shear compliance is a sum of the individual elements' shear compliances. Thus the model leads to the following expression, which  defines the parametrization used henceforth:

\be\label{NBO}
J(\omega) = J_\alpha\left( 1 + \frac{1}{i\omega \tau_\alpha} +
\frac{k_1}{1+k_2 (i\omega\tau_\alpha)^{1/2}}\right) +
\frac{J_\beta}{1 + (i\omega\tau_\beta)^b}\,.
\ee
For later use we note that the instantaneous (plateau) shear modulus $\Ginf=\lim_{\omega\rightarrow\infty}G(\omega)=\lim_{\omega\rightarrow\infty}1/J(\omega)$ is given by

\be\label{ginf_eq}
\Ginf
\,=\,\frac{1}{J_\alpha}\,.
\ee
The modulus plateau between the alpha and beta processes at temperatures low enough that these are well separated, i.e., when frequencies exist obeying $\omega\tau_\alpha \gg 1$ and $\omega\tau_\beta \ll 1$, is denoted by $G_{p,\alpha\beta}$ and given by 

\be\label{gp_eq}
G_{p,\alpha\beta}
\,=\,\frac{1}{J_\alpha+J_\beta}\,.
\ee
In the low-frequency limit the shear compliance diverges as $\propto 1/i\omega\tau_\alpha$ as required for any liquid with a finite DC viscosity. In the DC limit the real part of the shear compliance, the so-called recoverable shear compliance, is given by

\be\label{reccomp}
{\rm Re}\left(J(0)\right)
\,=\,(1+k_1)J_\alpha+J_\beta\,.
\ee

The model has seven parameters: 

\begin{itemize}
	\item two compliance strengths $J_\alpha$ and $J_\beta$ [unit: 1/GPa], 
	\item two relaxation times $\tau_\alpha$ and $\tau_\beta$ [unit: s],
	\item two alpha ``shape parameters'' $k_1$, $k_2$ [dimensionless],
	\item the beta CCRE exponent $b$ [dimensionless]. 
\end{itemize}
We shall assume that time-temperature superposition (TTS) applies for both the alpha and the beta processes, implying that in fit to data the three dimensionless shape parameters $k_1$, $k_2$, and $b$ cannot vary with temperature. The parameters allowed to vary with temperature are the two compliance strengths and the two relaxation times.

The characteristic times of the alpha and beta CPEs are denoted by $\tau_{c,\alpha}$ and $\tau_{c,\beta}$, respectively. In the below fit to squalane data we take as mentioned the constants $k_1$ and $k_2$ to be independent of temperature, which implies that $\tau_{c,\alpha}\propto\tau_{\alpha}$ with a temperature-independent constant of proportionality. For this reason $\tau_{c,\alpha}$ will not be discussed separately from $\tau_\alpha$. The beta characteristic time $\tau_{c,\beta}$, on the other hand, is not proportional to $\tau_\beta$ in its temperature variation, which makes both beta times important to keep track of (\sec{disc_sec}).

\section{Fitting the model to data for squalane}\label{fit_sec}

The model was fitted to the squalane data of \fig{G_data} using MATLAB's ``fminsearch'' Nelder-Mead downhill simplex least-squares fitting procedure. The fit excluded data taken at too low a temperature to be in equilibrium or at such high temperatures that the alpha and beta process have almost completely merged. These limitations leave data for temperatures between 168 K and 182 K for fitting and parameter identification. 

The data for the real and imaginary parts of the frequency-dependent shear modulus cover angular frequencies ranging from 10 mHz to 30 kHz, with up to 16 frequencies per decade evenly distributed on a logarithmic scale. The data were fitted to \eq{NBO} for the shear-modulus (fitting to the shear compliance would have been dominated by the low-frequency compliance divergence). First, the three temperature-independent shape parameters $k_1$, $k_2$, $b$ were identified by fitting to the 172 K data, which have the alpha and the beta loss peaks both well within the frequency window, but still clearly separated. Subsequently, the remaining four parameters were determined from the best fit at each temperature.

\begin{figure}[h!]
	\includegraphics[width=\textwidth]{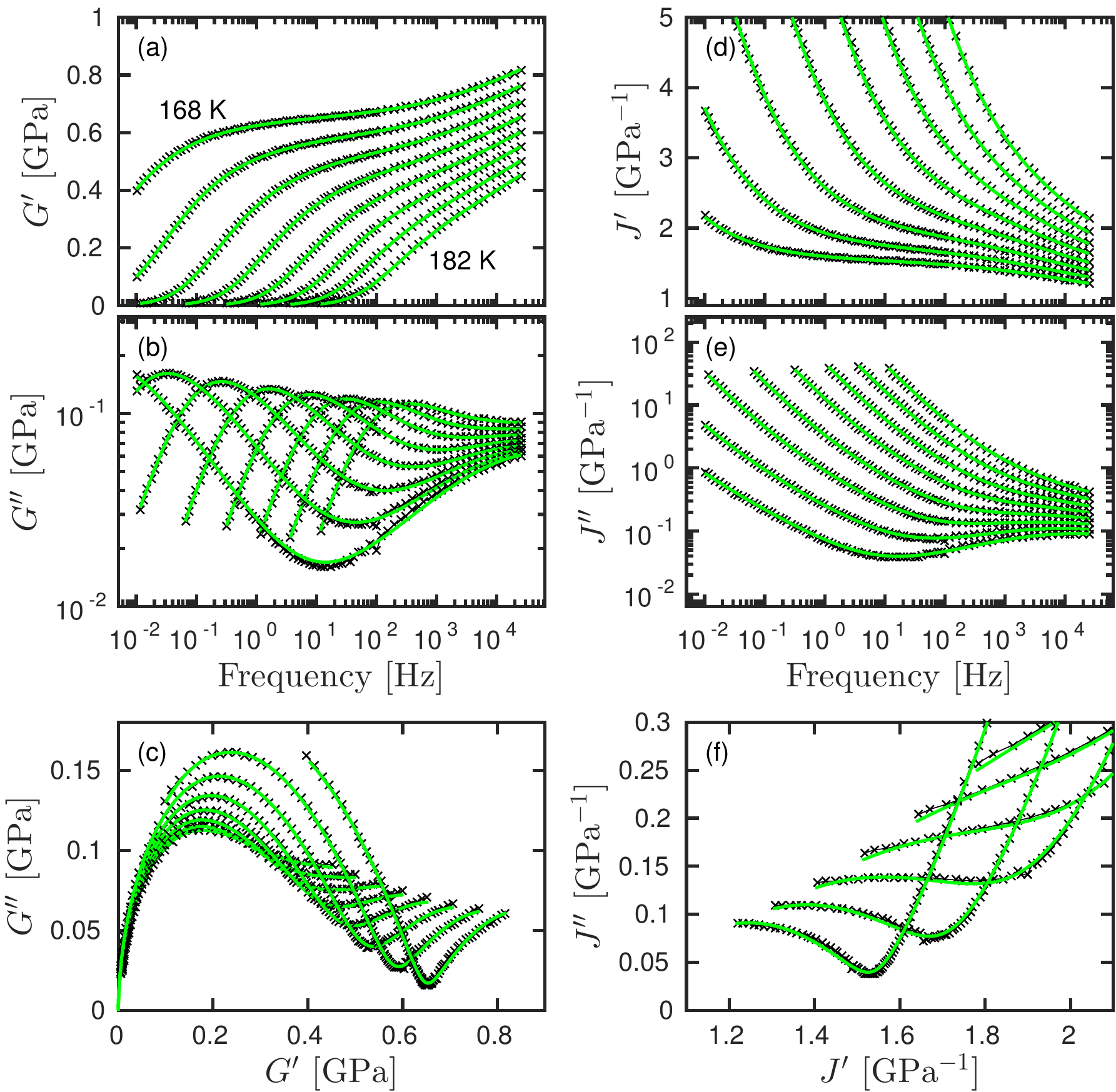}
	\caption{Comparison between data (black crosses) and model predictions \eq{NBO} (full green curves). At each temperature the four free parameters were determined as described in the text, but first the three temperature-independent shape parameters $k_1$, $k_2$, and $b$ were identified by fitting to the 172 K data. \label{NBO_vs_data}
	}
\end{figure}

Figure \ref{NBO_vs_data} compares model fits (full green curves) to data (black crosses). Figure \ref{NBO_vs_data}(a) gives model prediction versus data for the real part of the dynamic shear modulus, \fig{NBO_vs_data}(b) gives the same for the imaginary part, and \fig{NBO_vs_data}(c) gives model prediction versus data in a Nyquist plot of the shear modulus. Figures \ref{NBO_vs_data}(d), (e), and (f) give the same for the dynamic shear compliance.

\begin{figure}[h!]
	\includegraphics[width=8cm]{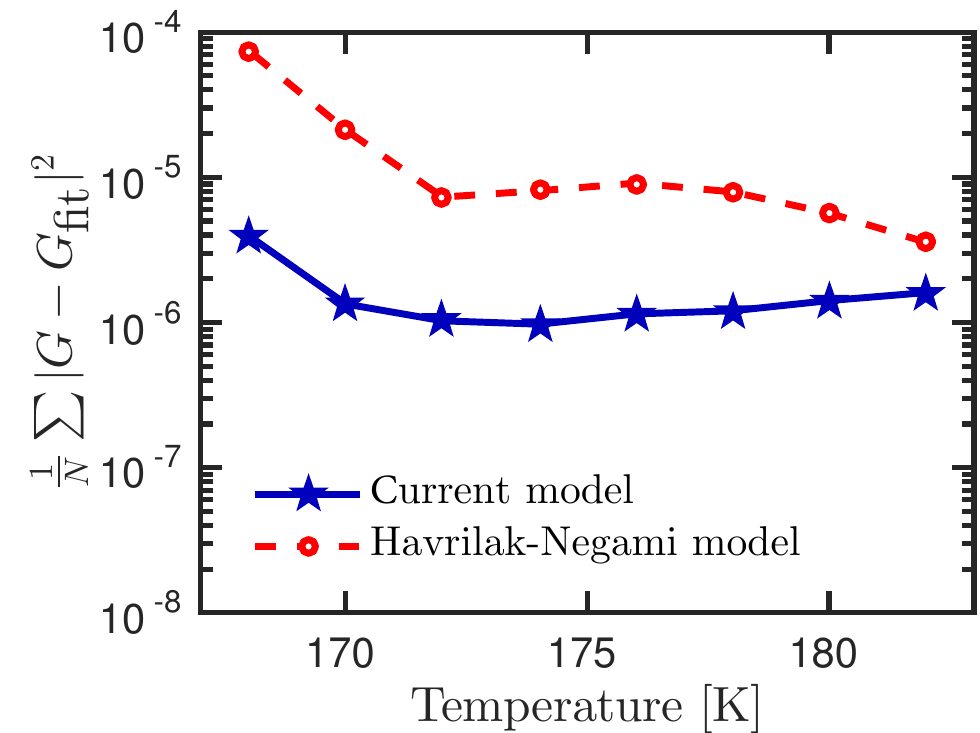}
	\caption{Comparing the quality of fitting data to \eq{NBO} and to \eq{HN}, quantified via the frequency-averaged deviation of the complex shear modulus from that of the data. Although the two models have the same number of parameters, the former fits data better by at least a factor of two and more than a factor of ten at low temperatures. \label{Fit_qual}}
\end{figure}

The fits are excellent, which given the number of free parameters may not appear very surprising. Our experience with fitting data to similar models over the last 20 years shows, however, that the present model is better than other models with the same number of parameters. As an illustration of this, we have compared to a fit assuming a Havriliak-Negami (HN) type function for the alpha process. The function fitted to data is the following: 

\begin{equation}\label{HN}
J(\omega) = J_\alpha^{\rm HN} \left( 1 - \frac{1}{(1+(i\omega \tau_\alpha^{\rm HN})^c)^a}\right)^{-1} +
\frac{J_\beta^{\rm HN}}{1 + (i\omega\tau_\beta^{\rm HN})^b}\,.
\end{equation}
This has the same number of parameters as \eq{NBO}: two strength parameters, two relaxation times, and three dimensionless shape parameters ($a,b,c$). A qualitative difference to \eq{NBO} should be mentioned, because the latter has a finite recoverable compliance, i.e., a finite low-frequency limit of $J'(\omega)$, whereas \eq{HN} like the BEL model \cite{BEL} diverges in this limit. We determined the best-fit parameters in the same way as above. The fit to data is not as good as that of \eq{NBO}, which is clear from \fig{Fit_qual} that compares the overall quality of the two best fits as functions of temperature.

\begin{figure}[h!]
  \includegraphics[width=8cm]{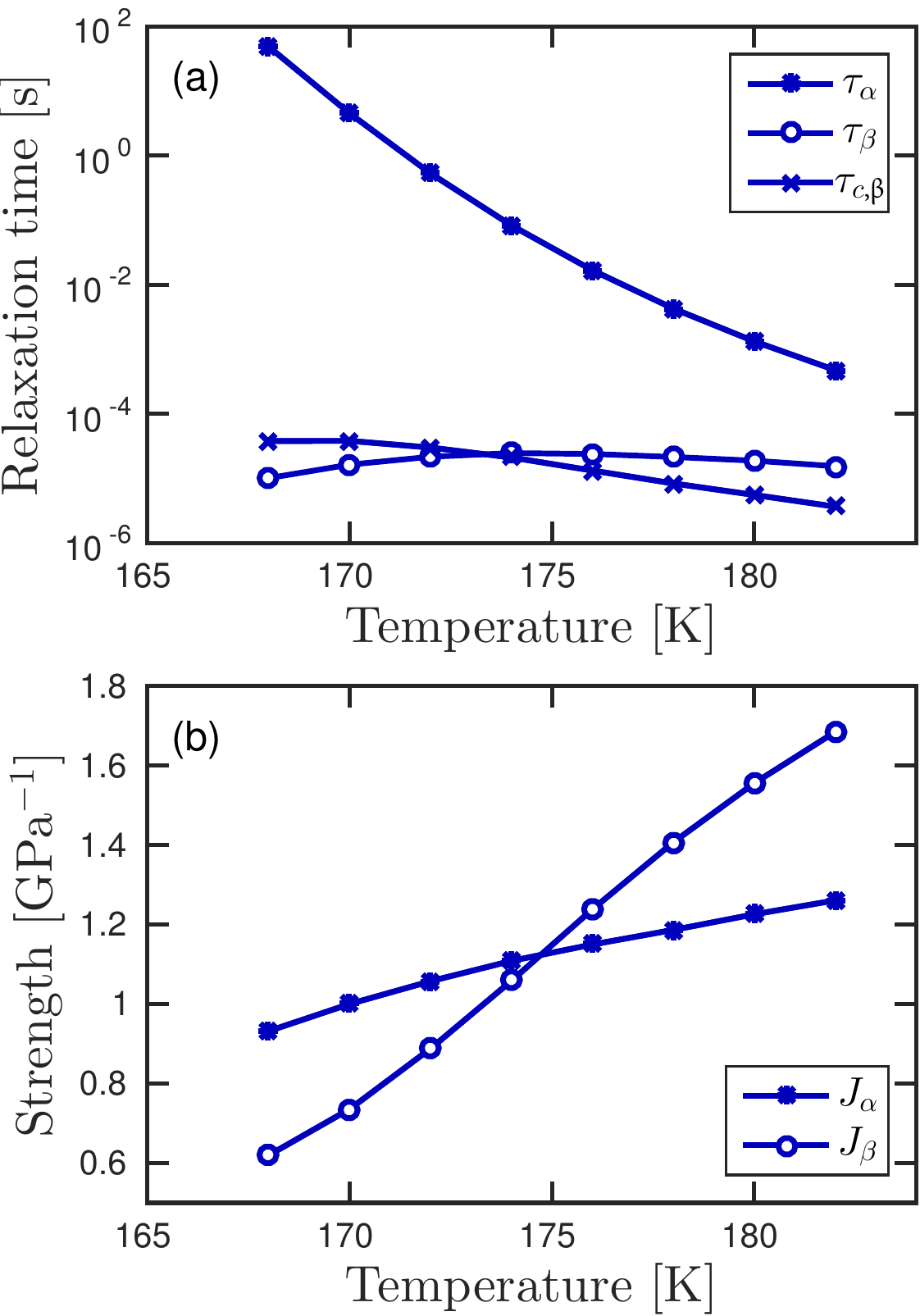}
  \caption{Temperature dependence of the parameters when \eq{NBO} is fitted to the shear-modulus data of \fig{G_data} assuming temperature independence of the three shape parameters $k_1$, $k_2$, and $b$, which were identified by fitting to the 172 K data resulting in $k_1=7.9$, $k_2=4.8$, $b=0.36$.
  (a) shows the alpha and beta relaxation times, $\tau_\alpha$ and $\tau_\beta$, as well as the beta characteristic time $\tau_{c,\beta}$. Note that $\tau_{c,\beta}$ differs from $\tau_\beta$ and has a more systematic variation with temperature. $\tau_\alpha$ decreases strongly with increasing temperature, which is a well-known feature of glass-forming liquids.
  (b) shows the temperature dependence of the shear-compliance strengths $J_\alpha$ and $J_\beta$. The latter changes by a almost factor of three whereas $J_\alpha$ changes by just 25\% over the same range of temperatures. The results of (a) and (b) for the beta process are qualitatively similar to previous findings of ours for the dielectric beta loss-peak frequency of glass-forming liquids, which is found to be Arrhenius in the glass phase but only weakly temperature dependent in the metastable liquid phase, whereas the beta strength has the opposite behavior with strong temperature dependence in the liquid and weak in the glass \cite{ols98a,ols00,dyr03}.\label{Param}
  }
\end{figure}

Returning to the model \eq{NBO}, Fig. \ref{Param} shows the temperature variation of the four free parameters. Figure \ref{Param}(a) shows how the alpha and beta relaxation times $\tau_\alpha$ and $\tau_\beta$ vary with temperature. As always for a glass-forming liquid, the alpha relaxation time increases strongly when temperature is decreased. The beta relaxation time $\tau_\beta$ is almost constant and, in fact, not even a monotonic function of temperature. In contrast, the beta \textit{characteristic} time $\tau_{c,\beta}$ decreases systematically with temperature. The dielectric beta loss-peak frequency is usually reported to be Arrhenius \cite{joh82,ric15}, but it is important to note that almost all literature data for $\tau_\beta$ (the inverse beta loss-peak frequency) refer to the glass phase, not to the metastable liquid phase about $T_g$. Figure \ref{Param}(b) shows the best-fit shear-compliance strengths $J_\alpha$ and $J_\beta$ as functions of temperature. Note that the beta process strength varies considerably more than the alpha strength.

These findings are consistent with previous ones for the beta dielectric relaxation process, which may be summarized as follows \cite{ols98a,ols00,dyr03,qia14a}: In the metastable liquid phase the relaxation strength increases considerably with increasing temperature whereas the relaxation time is almost temperature independent, in the glassy phase the strength is almost constant whereas the loss-peak frequency (inverse relaxation time) is strongly temperature dependent (Arrhenius). As an alternative to the minimal model of Ref. \onlinecite{dyr03} it is possible to rationalize these properties of the beta process -- as well as its behavior under annealing of the out-of-equilibrium liquid -- by assuming that the relaxation strength freezes at the glass transition whereas the characteristic time $\tau_{c,\beta}$ is Arrhenius with an activation energy that is unaffected by the glass transition (unpublished).

\section{Data for the dynamic adiabatic bulk modulus of squalane}\label{bulk_sec}

\begin{figure}[h!]
	\includegraphics[height=6cm]{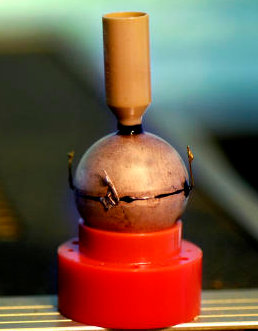}
	\caption{The piezo-ceramic transducer used for measuring the dynamic adiabatic bulk modulus \cite{chr94,iga08a,iga08b,hec13,gun14}. The device consists of a piezo-electric ceramic shell coated with electrodes on both sides with wires connecting the two electrodes to a frequency analyzer. The inner diameter is 18 mm, the shell thickness is 0.5 mm. A small hole in the shell makes it possible to fill the transducer with liquid at room temperature at which the viscosity is low. A tube acting as a liquid reservoir is attached on top of the hole, which ensures that the sphere remains filled as the liquid inside contracts upon cooling.		
	\label{PBG}}
\end{figure}

\begin{figure}
  \includegraphics[width=8cm]{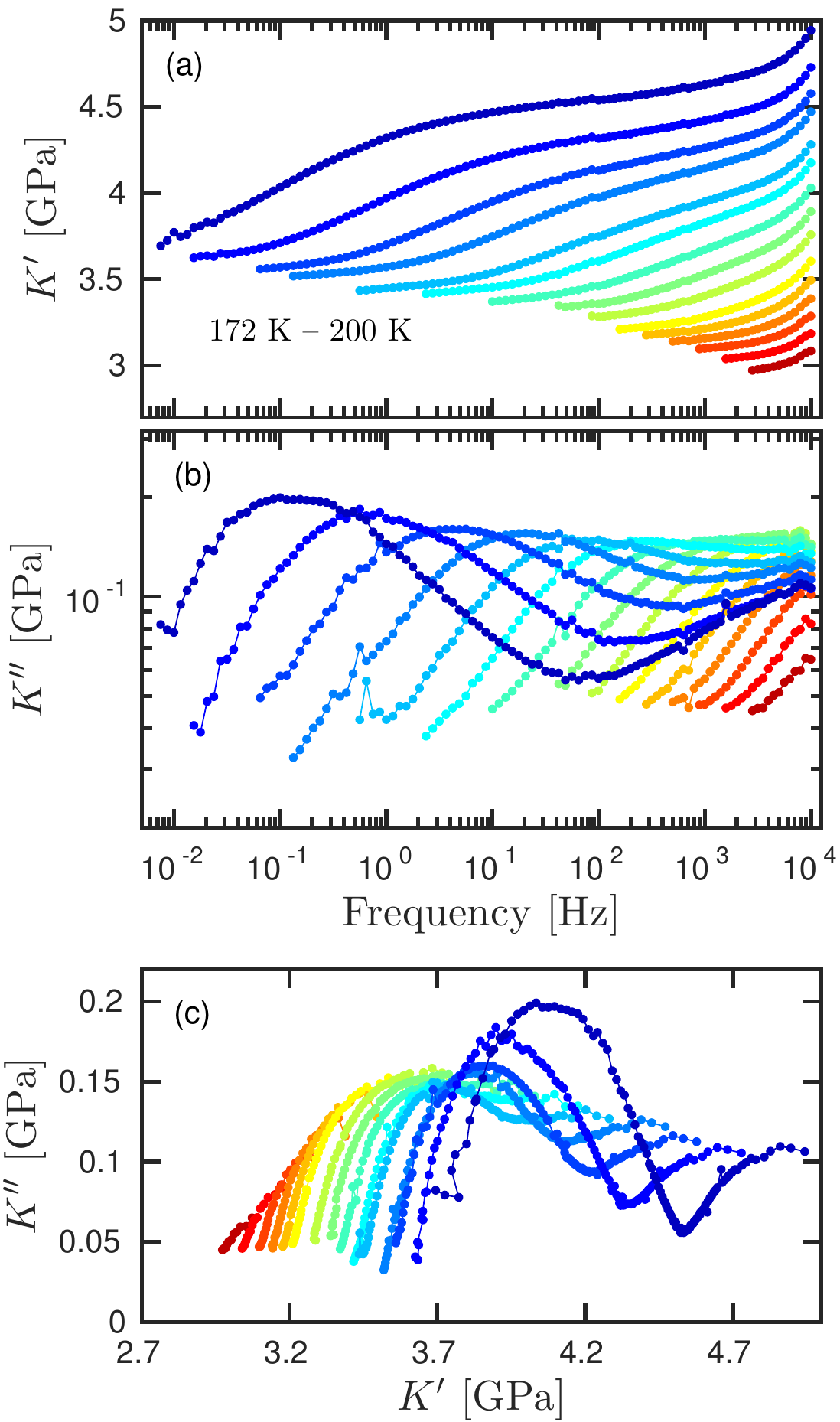}
  \caption{Data for the dynamic adiabatic bulk modulus of squalane $K(\omega)=K'(\omega)+iK''(\omega)$ covering temperatures from 172 K to 200 K.
  (a) and (b) show the real and the imaginary parts of $K(\omega)$, respectively.
  (c) shows a Nyquist plot of the same data.
  \label{K_data}}
\end{figure}

Figure \ref{PBG} shows our transducer for measuring the dynamic adiabatic bulk modulus. It consists of a radially polarized piezo-ceramic spherical shell coated with electrodes on the inner and outer surfaces. An applied electrical potential induces a slight compression or expansion of the sphere in which the liquid is placed (the top is a reservoir allowing for the liquid's thermal expansion) \cite{chr94,hec13,gun14}. Figure \ref{K_data} shows data for the real and imaginary parts of the dynamic adiabatic bulk modulus $K(\omega)=K'(\omega)+iK''(\omega)$, as well as a Nyquist plot of the same data.

Comparing the bulk modulus loss in \fig{K_data}(b) to the shear modulus loss in \fig{G_data}(b), we see a qualitatively similar behavior with an alpha loss peak that moves rapidly to lower frequencies upon cooling and a large beta peak appearing. To the best of our knowledge this is the first observation of a beta process for the dynamic bulk modulus. 

How to interpret the similarity between the dynamic shear and bulk moduli? This finding is certainly consistent with many previous ones \cite{jak12,hec13}, but it is important to emphasize that there is no fundamental reason for the similarity. This is because the dynamic bulk modulus -- whether adiabatic or isothermal -- is a \textit{scalar} linear-response function whereas the dynamic shear modulus is a \textit{vector} linear-response function. As discussed by Meixner long time ago these functions therefore belong to fundamentally different symmetry classes \cite{meixner}. Nevertheless, by reference to the Eshelby picture of structural rearrangements within a surrounding elastic matrix, Buchenau has recently discussed how the relaxational parts of the bulk and shear moduli may be connected \cite{buc12b} in arguments that may be extended to finite frequencies, thus establishing a connection between $G(\omega)$ and $K(\omega)$.

\begin{figure}[h!]
	\includegraphics[width=12cm]{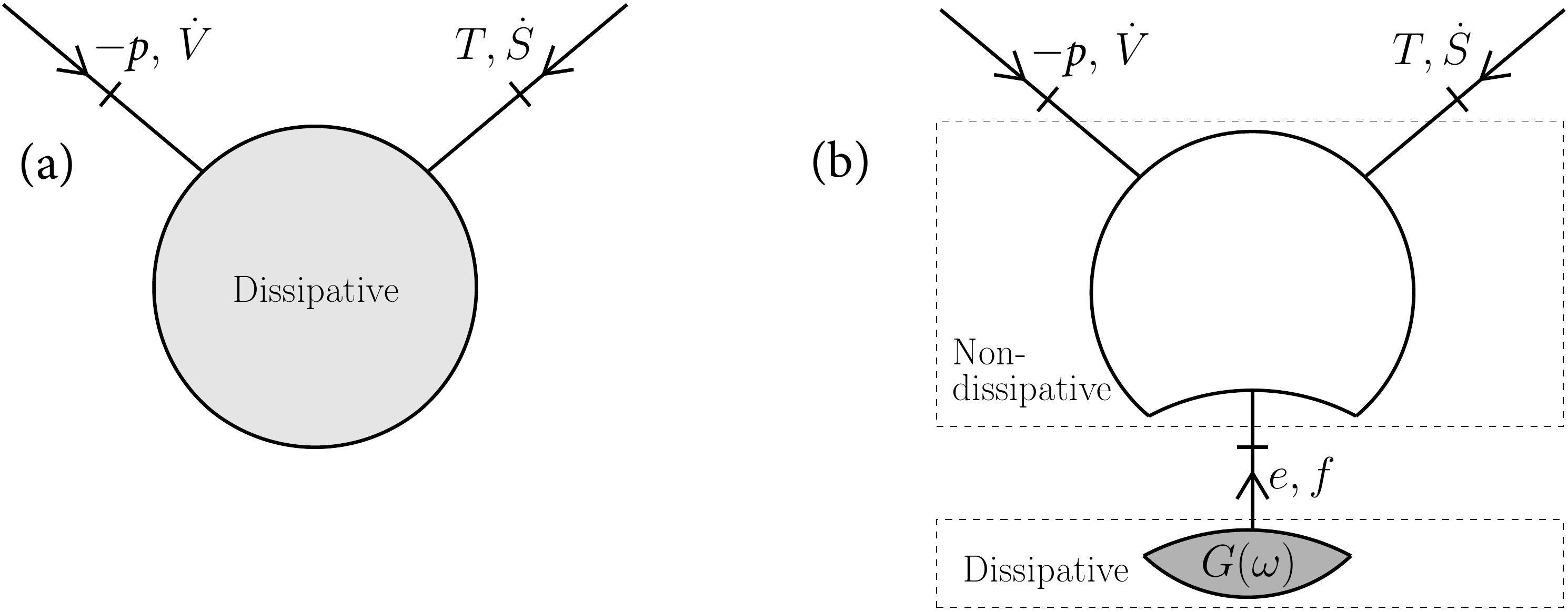}
	\caption{Schematic energy-bond diagram for the linear dynamic responses of the two fundamental scalar thermodynamic energy bonds: the thermal bond defined by effort and flow being temperature $T$ and entropy flow $\dot S$, respectively, and the mechanical bond defined by effort and flow being minus pressure $-p$ and rate of volume change $\dot V$, respectively. (a) The general scenario. (b) The situation in which all dissipation via a single ``internal'' energy bond is controlled by the dynamic shear modulus, an instance of the general single-order-parameter scenario \cite{ell07}. This diagram provides a link between the vector energy bond associated with mechanical shear deformation and the two scalar thermodynamic energy bonds. \label{One_param}}
\end{figure}

Referring to the energy-bond formalism \cite{paynter,ost71,pvc78,systemdyn,III}, there are two fundamental thermodynamic scalar energy bonds: a thermal energy bond with effort equal to temperature difference and flow equal to entropy current, and a mechanical bond with effort equal to minus pressure difference and flow equal to rate of volume change. Consistent with Buchenau's reasoning \cite{buc12b} we propose a general energy-bond model in which all dissipation connected with the two scalar thermodynamic energy bonds is controlled by the dynamic shear modulus (or, equivalently, the dynamic shear compliance). A representation of this idea is given in \fig{One_param}(b). An energy-bond diagram of this sort implies that the system in question is a ``single-order-parameter'' liquid \cite{ell07,bai08,III,gun11}. This is equivalent to being an R simple system, i.e., one with so-called isomorphs, which are lines in the thermodynamic phase diagram along which the dynamics is invariant to a very good approximation \cite{ped11,ing12,dyr14,dyr16}. 

In \fig{One_param}(b) there may be several non-dissipative elements, but the important point is that these are all connected to the element of  \fig{CIR2} via a single, internal energy bond. The predictions for the dynamic adiabatic/isothermal bulk moduli (or those of the dynamic expansion coefficient \cite{nis12}) depend, of course, not just on the dynamic shear modulus (compliance), but also on the non-dissipative elements. For a system described by \fig{One_param}(b) one \textit{a priori} expects that all the scalar response functions at any given temperature have alpha and beta processes located at frequencies similar to those of the shear modulus' alpha and beta processes.

\begin{figure}[h!]
	\includegraphics[width=7cm]{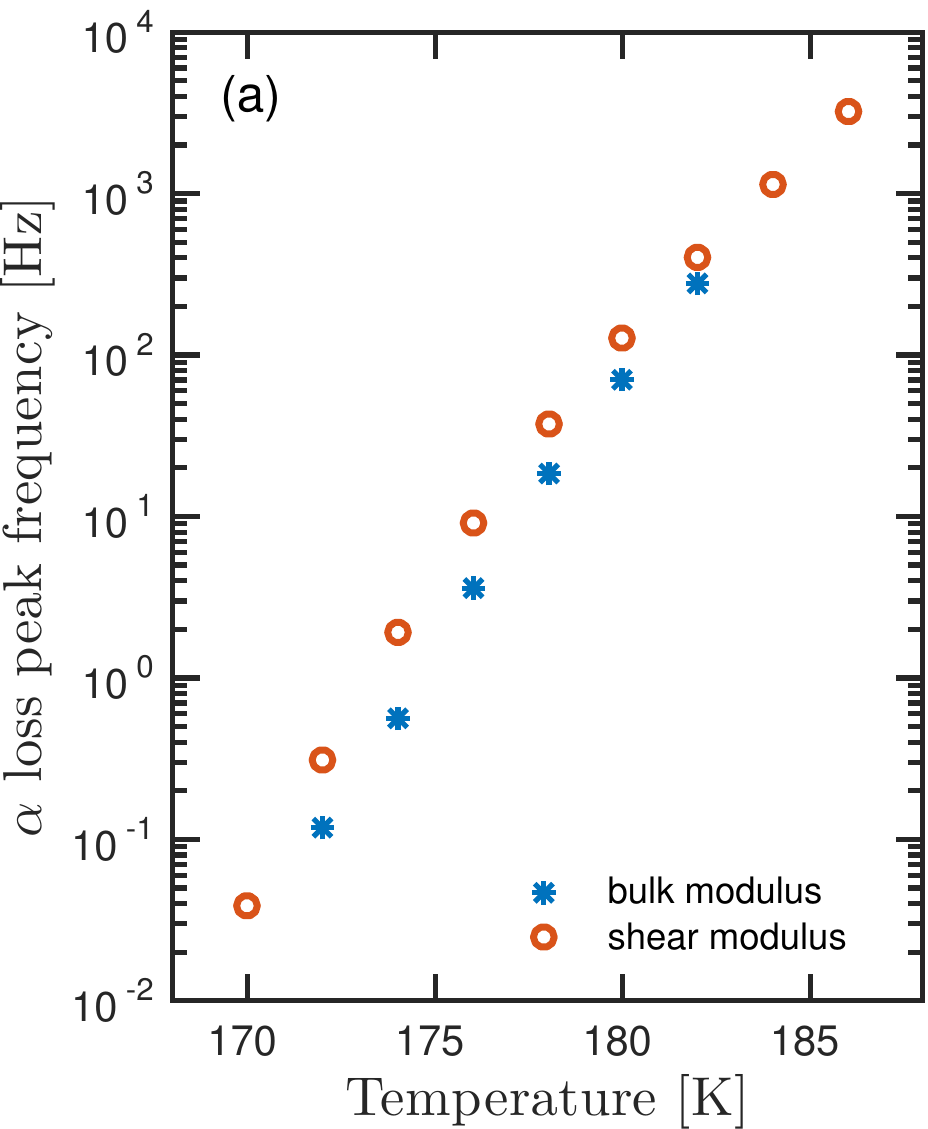}
	\includegraphics[width=7cm]{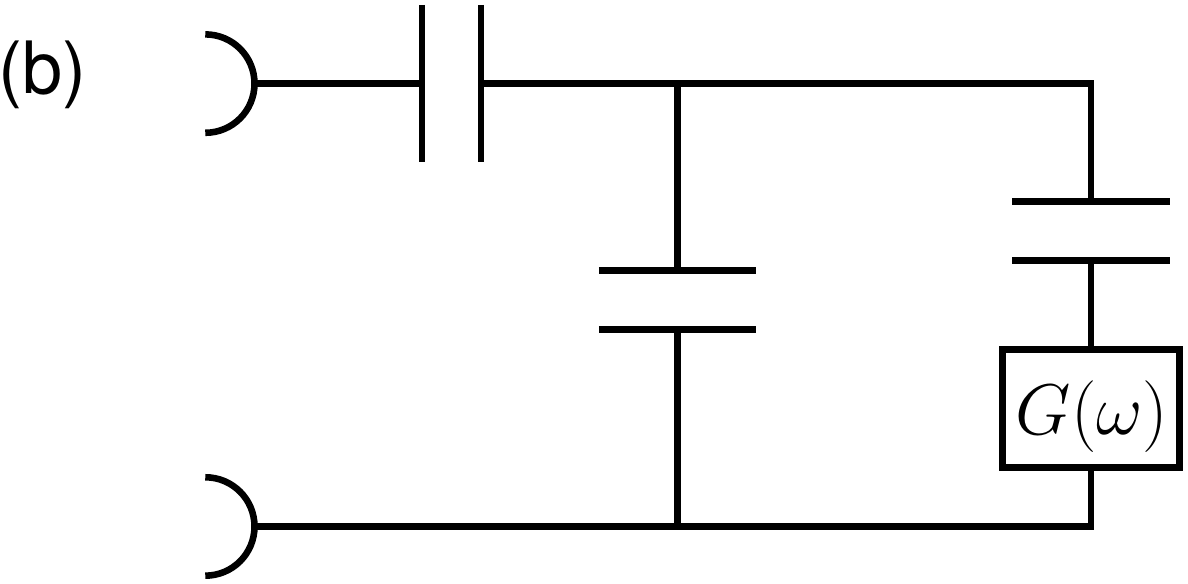}
	\includegraphics[width=7cm]{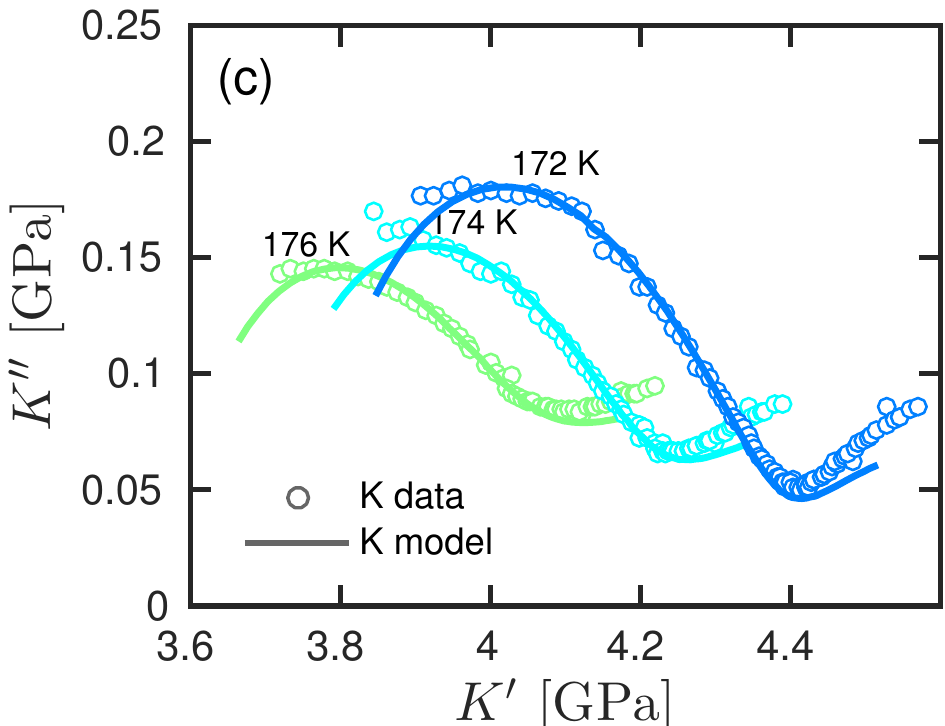}
	\caption{
	(a) Loss-peak frequencies of the dynamic shear modulus (red circles) and the dynamic adiabatic bulk modulus (blue asterices) as functions of temperature. Clearly the two probes of the dynamics have similar relaxation times and follow each other in the slowing down upon cooling. This may be rationalized by a model in which the bulk modulus is controlled by the shear modulus. 
	(b) An example of such a model. This is an instance of the general model philosophy illustrated in \fig{One_param}(b) in which all dissipation is controlled by the dynamic shear modulus.
	(c) Results from fitting the model in (b) to the dynamic bulk modulus data plotted in a Nyquist plot. Inputs to the fits at each temperature are the measured $G(\omega)$ and zero-frequency adiabatic bulk modulus $K(0)$, leaving two fitting parameters to determine the three capacitors of (b). We conclude that it is possible to fit the dynamic bulk modulus data in this way, at least qualitatively.
	\label{K_model} 	}
\end{figure}

As an example of the general modeling philosophy of \fig{One_param}(b), \fig{K_model}(b) gives a specific model for $K(\omega)$ in terms of $G(\omega)$. First, \fig{K_model}(a) demonstrates the similarity between the relaxation times of the equilibrium shear stress fluctuations determining $G(\omega)$ via the fluctuation-dissipation theorem (red circles) and those of the pressure fluctuations determining $K(\omega)$ (blue stars). Clearly these two times are of the same order of magnitude and have similar dependence on temperature. This means that a model following \fig{One_param}(b) makes sense. One may think of different such models, and the one shown in \fig{K_model}(b) is just an example. At each temperature there are only few fitting parameters while the entire frequency dependence and, in particular, the dissipation is determined by $G(\omega)$. In the fit to data we took the zero-frequency (adiabatic) bulk modulus $K(0)$ measured at the temperature in question as input, leaving just two free parameters. Nevertheless, the Nyquist plot of $K(\omega)$ demonstrates a reasonable fit (\fig{K_model}(c)).

\section{Discussion}\label{disc_sec}

This paper has demonstrated that dynamic shear-mechanical data for squalane may be fitted very well with the electrical-equivalent circuit model of \fig{CIR2} leading to \eq{NBO} for the shear compliance. The model assumes additivity of the alpha and beta shear compliances. The model has seven parameters, one more than alternative  phenomenological models \cite{jak11,buc16}. In the fit to data, however, the three dimensionless shape parameters were taken to be temperature independent, reflecting the assumption that time-temperature superposition applies separately to both the alpha and the beta compliance functions. In this picture, observed deviations from TTS derive from the merging of the alpha and beta processes. We conjecture that this applies generally for glass-forming liquids.

How to physically justify that the Maxwell RC element and the two CCREs should be combined in a way that is additive in their shear compliances, not in their shear moduli? There are no logically compelling arguments for this. We think of it as follows. Imagine a small particle in the liquid. The particle's mean-square displacement (MSD) as a function of time in one axis direction, $\langle\Delta x^2(t)\rangle$, will have a rapid increase on the phonon time scale, followed by a transition to the long-time diffusive behavior proportional to time. If one assumes that the alpha and beta processes are statistically independent \cite{die99}, this implies for the particle's motion that $\Delta x(t)=\Delta x_\alpha(t)+\Delta x_\beta(t)$ with $\langle\Delta x_\alpha(t)\Delta x_\beta(t)\rangle=0$. In this case the MSD is a sum of an alpha and a beta contribution:  $\langle\Delta x^2(t)\rangle =\langle\Delta x_\alpha^2(t)\rangle +\langle\Delta x_\beta^2(t)\rangle$. If one moreover assumes the Stokes-Einstein relation between the dynamic shear viscosity and the particle's dynamic friction coefficient, this translates via the fluctuation-dissipation theorem into additivity of the dynamic shear compliances for the alpha and beta processes. 

In regard to the single-particle MSD, note that associated with any function $\langle\Delta x^2(t)\rangle$ there is a characteristic time, namely the time at which the particle has moved a typical intermolecular distance. This is how we think of each CPE basic element's characteristic $\tau_c$, which was defined by the absolute value of the compliance at $\omega=1/\tau_c$ being 1 ${\rm GPa}^{-1}$ (\eq{gpainv}). For the beta process, it is important to distinguish between this time and the inverse loss-peak frequency $\tau_\beta$, because via \eq{taueq} the latter time's temperature variation reflects the combined effect of the changing compliance strength $J_\beta$ and the Arrhenius $\tau_{c,\beta}$. 

The electrical-equivalent circuit model \fig{CIR2} is identical to that proposed in Ref. \onlinecite{jak11} except for an extra capacitor, the one in the alpha CCRE. This capacitor eliminates an unphysical feature of our previous model \cite{jak11}, which predicted an infinite recoverable shear compliance. This unphysical feature is also present in the BEL model from 1967 \cite{BEL,har76}. Introducing the extra capacitor has the added benefit of resulting in symmetry between the alpha and beta CCREs, the only difference being that the alpha CCRE has the exponent fixed to $1/2$. 

We note that if the alpha CCRE's relaxation time is much shorter than $RC$, the circuit mimics the situation reported in recent papers for the dielectric relaxation of monohydroxy alcohols, for which one observes a low-frequency Debye-type process followed by, in order of increasing frequency, first an alpha and then, in most liquids, a beta process \cite{boh14a,gai14}.

\begin{figure}[h!]
	\includegraphics[width=8cm]{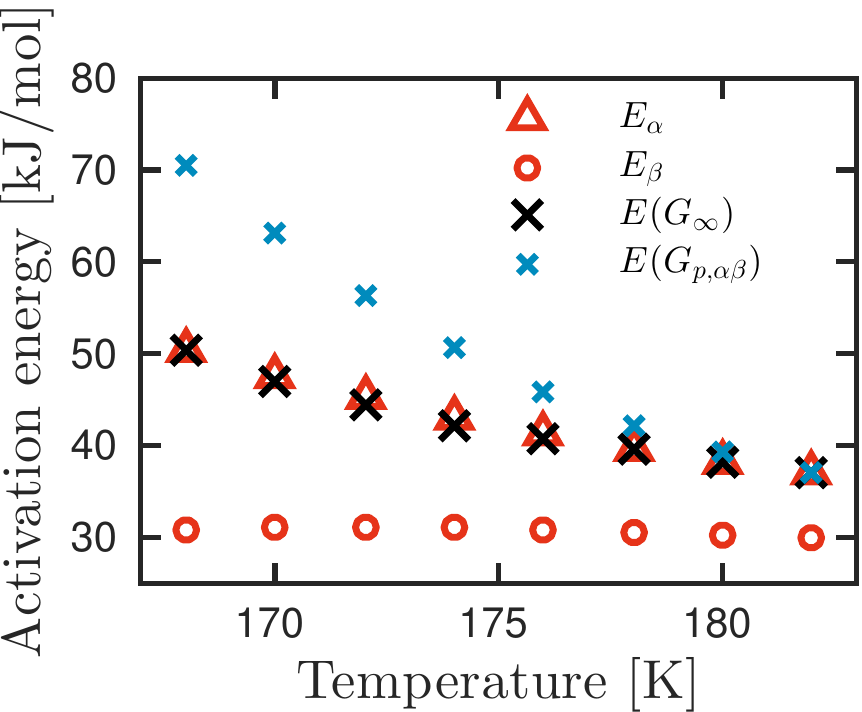}
	\caption{Activation energies calculated from $E=RT\ln(\tau/\tau_0)$ for the alpha relaxation time $\tau_\alpha$ (red triangles) and the beta characteristic time $\tau_{c,\beta}$ (red circles), assuming in both cases that $\tau_0=10^{-14}$. $E_\beta$ is temperature independent to a good approximation, showing that $\tau_{c,\beta}$ is Arrhenius. The alpha relaxation time activation energy is compared to the prediction of the shoving model (black crosses) according to which $E_\alpha=V_cG_\infty$ where $V_c$ is the ``characteristic'' (temperature-independent) volume and $G_\infty$ is the instantaneous, i.e., high-frequency limiting shear modulus, which according to \eq{NBO} equals $1/J_\alpha$. The shoving model is confirmed. If one in this model, however, instead of $G_\infty$ uses the modulus corresponding to frequencies between the alpha and beta process, $G_{p,\alpha\beta}$ (\eq{gp_eq}), the $E_\alpha$ prediction results in the blue crosses that do not fit the red triangle data. The shoving-model based activation energies $E(G_\infty)$ and $E(G_{p,\alpha\beta})$ were both normalized to predict the correct alpha activation energy at $T=182$ K (leading to $V_c$ being 9\% of the molar volume for the $E(G_\infty)$ case).		\label{activE} 	}
\end{figure}

As regards the temperature dependence of the best-fit model parameters for squalane we have compared to the prediction of the shoving model \cite{dyr96,dyr06}. If $\tau_{0}\cong 10^{-14}$s is a typical phonon time and $V_c$ the so-called characteristic volume assumed to be temperature independent, the shoving model predicts the following relation between the temperature variation of the alpha relaxation time and that of the instantaneous, i.e., high-frequency plateau, shear modulus $G_\infty$:

\be\label{shov}
\tau_\alpha (T)
\,=\,\tau_{0}\, e^{V_c\Ginf(T)/k_BT}\,.
\ee
The shoving model, which links a supercooled liquid's fragility to the temperature variation of $\Ginf$, fits data well for many glass-forming liquids \cite{dyr96,dyr06,hec13,hec15a}. The shoving model relates directly to \eq{NBO} since if it applies, via \eq{ginf_eq} the number of temperature-dependent parameters is reduced from four to three. We have not made this assumption in the fit to data, but have instead checked \eq{shov} against the best-fit parameters. This is done in \fig{activE} in which the relaxation times $\tau_\alpha$ and $\tau_{c,\beta}$ have been converted into temperature-dependent activation energies $E(T)$ by writing for each $\tau(T)=\tau_{0}\exp(E(T)/RT)$ with $\tau_{0}=10^{-14}$s. According to the shoving model $E_\alpha(T)=V_cG_\infty(T)$. By comparing the black crosses and the red triangles in \fig{activE} we conclude that the shoving model applies with $G_\infty$ calculated from the best fit model parameters via \eq{ginf_eq}. 

If the shoving model is instead interpreted with $G_\infty$ taken to be the modulus between the alpha and beta relaxations, the quantity $G_{p,\alpha\beta}$ given by \eq{gp_eq}, the model does not apply (blue crosses). This confirms the basic physical assumption of the shoving model, which is that the actual barrier transition for a rearrangement of molecules is very fast, presumably on the picosecond time scale. Consequently, the activation energy is proportional to the shear modulus of the liquid measured on this short time scale (at which the liquid behaves like a solid), not to the plateau modulus between the alpha and beta relaxations.

Our new model for $G(\omega)$ (\fig{CIR2}) consists of a Maxwell element in parallel with two CCREs. One may speculate that additional high-frequency mechanical processes beyond the alpha and beta relaxations can be included by adding further CCREs in parallel, each one still subject to time-temperature superposition, i.e., with temperature-independent shape parameters.

To summarize, an excellent fit to dynamic shear-mechanical data of squalane is provided by an electrical equivalent circuit model with seven parameters. The model assumes an $\omega^{-1/2}$ high-frequency decay of the alpha compliance \cite{BEL,dyr05,dyr06a,dyr07}, additivity of the alpha and beta compliance functions, and that these functions separately obey time-temperature superposition. The latter assumption reduces the number of parameters varying with temperature to four. The best fit parameters confirm the shoving model and show that the beta process characteristic time has a temperature-independent activation energy. If these findings were both incorporated as model assumptions, the number of parameters varying with temperature reduces to two. These could be taken to be, e.g., the compliance magnitudes $J_\alpha$ and $J_\beta$. We also presented data for the adiabatic dynamic bulk modulus and showed that these may be interpreted in terms of a single-order-parameter model in which all dissipation is controlled by the shear-mechanical properties. Such a model connects the class of scalar viscoelastic linear-response functions to that of vector symmetry \cite{meixner}.

In regard to future works, one obvious thing is to compare the model to shear-mechanical data for other glass-forming liquids. We have not done so systematically, but have found in all cases tested so far that the model works well (unpublished). The hope is that the model is general, thus providing a step towards a microscopic understanding of supercooled liquids' shear-dynamical properties.

\acknowledgments{Figure \ref{One_param}(b) is a special case of the general energy-bond diagram for a system with a single order parameter identified by our \textit{Glass and Time} colleague Tage Christensen about 15 years ago, which initiated a development leading to Ref. \onlinecite{ell07} that in turn led to the isomorph theory \cite{ped08,ing12,dyr14,dyr16}.}


\begin{thebibliography}{84}%
\makeatletter
\providecommand \@ifxundefined [1]{%
 \@ifx{#1\undefined}
}%
\providecommand \@ifnum [1]{%
 \ifnum #1\expandafter \@firstoftwo
 \else \expandafter \@secondoftwo
 \fi
}%
\providecommand \@ifx [1]{%
 \ifx #1\expandafter \@firstoftwo
 \else \expandafter \@secondoftwo
 \fi
}%
\providecommand \natexlab [1]{#1}%
\providecommand \enquote  [1]{``#1''}%
\providecommand \bibnamefont  [1]{#1}%
\providecommand \bibfnamefont [1]{#1}%
\providecommand \citenamefont [1]{#1}%
\providecommand \href@noop [0]{\@secondoftwo}%
\providecommand \href [0]{\begingroup \@sanitize@url \@href}%
\providecommand \@href[1]{\@@startlink{#1}\@@href}%
\providecommand \@@href[1]{\endgroup#1\@@endlink}%
\providecommand \@sanitize@url [0]{\catcode `\\12\catcode `\$12\catcode
  `\&12\catcode `\#12\catcode `\^12\catcode `\_12\catcode `\%12\relax}%
\providecommand \@@startlink[1]{}%
\providecommand \@@endlink[0]{}%
\providecommand \url  [0]{\begingroup\@sanitize@url \@url }%
\providecommand \@url [1]{\endgroup\@href {#1}{\urlprefix }}%
\providecommand \urlprefix  [0]{URL }%
\providecommand \Eprint [0]{\href }%
\providecommand \doibase [0]{http://dx.doi.org/}%
\providecommand \selectlanguage [0]{\@gobble}%
\providecommand \bibinfo  [0]{\@secondoftwo}%
\providecommand \bibfield  [0]{\@secondoftwo}%
\providecommand \translation [1]{[#1]}%
\providecommand \BibitemOpen [0]{}%
\providecommand \bibitemStop [0]{}%
\providecommand \bibitemNoStop [0]{.\EOS\space}%
\providecommand \EOS [0]{\spacefactor3000\relax}%
\providecommand \BibitemShut  [1]{\csname bibitem#1\endcsname}%
\let\auto@bib@innerbib\@empty
\bibitem [{\citenamefont {Harrison}(1976)}]{har76}%
  \BibitemOpen
  \bibfield  {author} {\bibinfo {author} {\bibfnamefont {G.}~\bibnamefont
  {Harrison}},\ }\href@noop {} {\emph {\bibinfo {title} {The Dynamic Properties
  of Supercooled Liquids}}}\ (\bibinfo  {publisher} {Academic (New York)},\
  \bibinfo {year} {1976})\BibitemShut {NoStop}%
\bibitem [{\citenamefont {Brawer}(1985)}]{bra85}%
  \BibitemOpen
  \bibfield  {author} {\bibinfo {author} {\bibfnamefont {S.}~\bibnamefont
  {Brawer}},\ }\href@noop {} {\emph {\bibinfo {title} {Relaxation in Viscous
  Liquids and Glasses}}}\ (\bibinfo  {publisher} {American Ceramic Society,
  Columbus, OH},\ \bibinfo {year} {1985})\BibitemShut {NoStop}%
\bibitem [{\citenamefont {Ediger}\ \emph {et~al.}(1996)\citenamefont {Ediger},
  \citenamefont {Angell},\ and\ \citenamefont {Nagel}}]{edi96}%
  \BibitemOpen
  \bibfield  {author} {\bibinfo {author} {\bibfnamefont {M.~D.}\ \bibnamefont
  {Ediger}}, \bibinfo {author} {\bibfnamefont {C.~A.}\ \bibnamefont {Angell}},
  \ and\ \bibinfo {author} {\bibfnamefont {S.~R.}\ \bibnamefont {Nagel}},\
  }\bibfield  {title} {\enquote {\bibinfo {title} {Supercooled liquids and
  glasses},}\ }\href@noop {} {\bibfield  {journal} {\bibinfo  {journal} {J.
  Phys. Chem}\ }\textbf {\bibinfo {volume} {100}},\ \bibinfo {pages} {13200}
  (\bibinfo {year} {1996})}\BibitemShut {NoStop}%
\bibitem [{\citenamefont {Angell}(1995)}]{ang95}%
  \BibitemOpen
  \bibfield  {author} {\bibinfo {author} {\bibfnamefont {C.~A.}\ \bibnamefont
  {Angell}},\ }\bibfield  {title} {\enquote {\bibinfo {title} {Formation of
  glasses from liquids and biopolymers},}\ }\href@noop {} {\bibfield  {journal}
  {\bibinfo  {journal} {Science}\ }\textbf {\bibinfo {volume} {267}},\ \bibinfo
  {pages} {1924--1935} (\bibinfo {year} {1995})}\BibitemShut {NoStop}%
\bibitem [{\citenamefont {Dyre}(2006{\natexlab{a}})}]{dyr06}%
  \BibitemOpen
  \bibfield  {author} {\bibinfo {author} {\bibfnamefont {J.~C.}\ \bibnamefont
  {Dyre}},\ }\bibfield  {title} {\enquote {\bibinfo {title} {{The Glass
  Transition and Elastic Models of Glass-Forming Liquids}},}\ }\href@noop {}
  {\bibfield  {journal} {\bibinfo  {journal} {Rev. Mod. Phys.}\ }\textbf
  {\bibinfo {volume} {78}},\ \bibinfo {pages} {953--972} (\bibinfo {year}
  {2006}{\natexlab{a}})}\BibitemShut {NoStop}%
\bibitem [{\citenamefont {Berthier}\ and\ \citenamefont
  {Biroli}(2011)}]{ber11}%
  \BibitemOpen
  \bibfield  {author} {\bibinfo {author} {\bibfnamefont {L.}~\bibnamefont
  {Berthier}}\ and\ \bibinfo {author} {\bibfnamefont {G.}~\bibnamefont
  {Biroli}},\ }\bibfield  {title} {\enquote {\bibinfo {title} {{Theoretical
  Perspective on the Glass Transition and Amorphous Materials}},}\ }\href@noop
  {} {\bibfield  {journal} {\bibinfo  {journal} {Rev. Mod. Phys.}\ }\textbf
  {\bibinfo {volume} {83}},\ \bibinfo {pages} {587--645} (\bibinfo {year}
  {2011})}\BibitemShut {NoStop}%
\bibitem [{\citenamefont {Floudas}\ \emph {et~al.}(2011)\citenamefont
  {Floudas}, \citenamefont {Paluch}, \citenamefont {Grzybowski},\ and\
  \citenamefont {Ngai}}]{flo11}%
  \BibitemOpen
  \bibfield  {author} {\bibinfo {author} {\bibfnamefont {G.}~\bibnamefont
  {Floudas}}, \bibinfo {author} {\bibfnamefont {M.}~\bibnamefont {Paluch}},
  \bibinfo {author} {\bibfnamefont {A.}~\bibnamefont {Grzybowski}}, \ and\
  \bibinfo {author} {\bibfnamefont {K.L.}\ \bibnamefont {Ngai}},\ }\href@noop
  {} {\emph {\bibinfo {title} {{Molecular Dynamics of Glass-Forming Systems:
  Effects of Pressure}}}}\ (\bibinfo  {publisher} {Springer, Berlin},\ \bibinfo
  {year} {2011})\BibitemShut {NoStop}%
\bibitem [{\citenamefont {Stillinger}\ and\ \citenamefont
  {Debenedetti}(2013)}]{sti13}%
  \BibitemOpen
  \bibfield  {author} {\bibinfo {author} {\bibfnamefont {F.~H.}\ \bibnamefont
  {Stillinger}}\ and\ \bibinfo {author} {\bibfnamefont {P.~G.}\ \bibnamefont
  {Debenedetti}},\ }\bibfield  {title} {\enquote {\bibinfo {title} {Glass
  transition -- thermodynamics and kinetics},}\ }\href@noop {} {\bibfield
  {journal} {\bibinfo  {journal} {Annu. Rev. Condens. Matter Phys.}\ }\textbf
  {\bibinfo {volume} {4}},\ \bibinfo {pages} {263--285} (\bibinfo {year}
  {2013})}\BibitemShut {NoStop}%
\bibitem [{\citenamefont {Deegan}\ \emph {et~al.}(1999)\citenamefont {Deegan},
  \citenamefont {Leheny}, \citenamefont {Menon}, \citenamefont {Nagel},\ and\
  \citenamefont {Venerus}}]{dee99}%
  \BibitemOpen
  \bibfield  {author} {\bibinfo {author} {\bibfnamefont {R.~D.}\ \bibnamefont
  {Deegan}}, \bibinfo {author} {\bibfnamefont {R.~L.}\ \bibnamefont {Leheny}},
  \bibinfo {author} {\bibfnamefont {N.}~\bibnamefont {Menon}}, \bibinfo
  {author} {\bibfnamefont {S.~R.}\ \bibnamefont {Nagel}}, \ and\ \bibinfo
  {author} {\bibfnamefont {D.~C.}\ \bibnamefont {Venerus}},\ }\bibfield
  {title} {\enquote {\bibinfo {title} {Dynamic shear modulus of tricresyl
  phosphate and squalane},}\ }\href {\doibase 10.1021/jp983832g} {\bibfield
  {journal} {\bibinfo  {journal} {J. Phys. Chem. B}\ }\textbf {\bibinfo
  {volume} {103}},\ \bibinfo {pages} {4066--4070} (\bibinfo {year}
  {1999})}\BibitemShut {NoStop}%
\bibitem [{\citenamefont {Richert}\ \emph {et~al.}(2003)\citenamefont
  {Richert}, \citenamefont {Duvvuri},\ and\ \citenamefont {Duong}}]{ric03}%
  \BibitemOpen
  \bibfield  {author} {\bibinfo {author} {\bibfnamefont {R}~\bibnamefont
  {Richert}}, \bibinfo {author} {\bibfnamefont {K}~\bibnamefont {Duvvuri}}, \
  and\ \bibinfo {author} {\bibfnamefont {LT}~\bibnamefont {Duong}},\ }\bibfield
   {title} {\enquote {\bibinfo {title} {Dynamics of glass-forming liquids.
  {VII.} {Dielectric} relaxation of supercooled tris-naphthylbenzene, squalane,
  and decahydroisoquinoline},}\ }\href {\doibase 10.1063/1.1531587} {\bibfield
  {journal} {\bibinfo  {journal} {J. Chem. Phys.}\ }\textbf {\bibinfo {volume}
  {118}},\ \bibinfo {pages} {1828--1836} (\bibinfo {year} {2003})}\BibitemShut
  {NoStop}%
\bibitem [{\citenamefont {Jakobsen}\ \emph {et~al.}(2005)\citenamefont
  {Jakobsen}, \citenamefont {Niss},\ and\ \citenamefont {Olsen}}]{jak05}%
  \BibitemOpen
  \bibfield  {author} {\bibinfo {author} {\bibfnamefont {B.}~\bibnamefont
  {Jakobsen}}, \bibinfo {author} {\bibfnamefont {K.}~\bibnamefont {Niss}}, \
  and\ \bibinfo {author} {\bibfnamefont {N.~B.}\ \bibnamefont {Olsen}},\
  }\bibfield  {title} {\enquote {\bibinfo {title} {Dielectric and shear
  mechanical alpha and beta relaxations in seven glass-forming liquids},}\
  }\href@noop {} {\bibfield  {journal} {\bibinfo  {journal} {J. Chem. Phys.}\
  }\textbf {\bibinfo {volume} {123}},\ \bibinfo {pages} {234511} (\bibinfo
  {year} {2005})}\BibitemShut {NoStop}%
\bibitem [{\citenamefont {Comunas}\ \emph {et~al.}(2014)\citenamefont
  {Comunas}, \citenamefont {Paredes}, \citenamefont {Gacino}, \citenamefont
  {Fernandez}, \citenamefont {Bazile}, \citenamefont {Boned}, \citenamefont
  {Daridon}, \citenamefont {Galliero}, \citenamefont {Pauly},\ and\
  \citenamefont {Harris}}]{com14}%
  \BibitemOpen
  \bibfield  {author} {\bibinfo {author} {\bibfnamefont {M.~J.~P.}\
  \bibnamefont {Comunas}}, \bibinfo {author} {\bibfnamefont {X.}~\bibnamefont
  {Paredes}}, \bibinfo {author} {\bibfnamefont {F.~M.}\ \bibnamefont {Gacino}},
  \bibinfo {author} {\bibfnamefont {J.}~\bibnamefont {Fernandez}}, \bibinfo
  {author} {\bibfnamefont {J.-P.}\ \bibnamefont {Bazile}}, \bibinfo {author}
  {\bibfnamefont {C.}~\bibnamefont {Boned}}, \bibinfo {author} {\bibfnamefont
  {J.~L.}\ \bibnamefont {Daridon}}, \bibinfo {author} {\bibfnamefont
  {G.}~\bibnamefont {Galliero}}, \bibinfo {author} {\bibfnamefont
  {J.}~\bibnamefont {Pauly}}, \ and\ \bibinfo {author} {\bibfnamefont {K.~R.}\
  \bibnamefont {Harris}},\ }\bibfield  {title} {\enquote {\bibinfo {title}
  {{Viscosity Measurements for Squalane at High Pressures to 350 MPa from T =
  (293.15 to 363.15) K}},}\ }\href@noop {} {\bibfield  {journal} {\bibinfo
  {journal} {J. Chem. Thermodyn.}\ }\textbf {\bibinfo {volume} {69}},\ \bibinfo
  {pages} {201--208} (\bibinfo {year} {2014})}\BibitemShut {NoStop}%
\bibitem [{\citenamefont {Angell}\ \emph {et~al.}(2000)\citenamefont {Angell},
  \citenamefont {Ngai}, \citenamefont {McKenna}, \citenamefont {McMillan},\
  and\ \citenamefont {Martin}}]{ang00}%
  \BibitemOpen
  \bibfield  {author} {\bibinfo {author} {\bibfnamefont {C.~A.}\ \bibnamefont
  {Angell}}, \bibinfo {author} {\bibfnamefont {K.~L.}\ \bibnamefont {Ngai}},
  \bibinfo {author} {\bibfnamefont {G.~B.}\ \bibnamefont {McKenna}}, \bibinfo
  {author} {\bibfnamefont {P.~F.}\ \bibnamefont {McMillan}}, \ and\ \bibinfo
  {author} {\bibfnamefont {S.~W.}\ \bibnamefont {Martin}},\ }\bibfield  {title}
  {\enquote {\bibinfo {title} {Relaxation in glassforming liquids and amorphous
  solids},}\ }\href@noop {} {\bibfield  {journal} {\bibinfo  {journal} {J.
  Appl. Phys.}\ }\textbf {\bibinfo {volume} {88}},\ \bibinfo {pages}
  {3113--3157} (\bibinfo {year} {2000})}\BibitemShut {NoStop}%
\bibitem [{wik(2017)}]{wiki}%
  \BibitemOpen
  \href@noop {} {\enquote {\bibinfo {title} {{Squalane}},}\ }\bibinfo
  {howpublished} {Wikipedia article} (\bibinfo {year} {2017})\BibitemShut
  {NoStop}%
\bibitem [{\citenamefont {Bair}(2006)}]{bai11}%
  \BibitemOpen
  \bibfield  {author} {\bibinfo {author} {\bibfnamefont {Scott}\ \bibnamefont
  {Bair}},\ }\bibfield  {title} {\enquote {\bibinfo {title} {Reference liquids
  for quantitative elastohydrodynamics: selection and rheological
  characterization},}\ }\href {\doibase 10.1007/s11249-006-9083-y} {\bibfield
  {journal} {\bibinfo  {journal} {Tribol. Lett.}\ }\textbf {\bibinfo {volume}
  {22}},\ \bibinfo {pages} {197--206} (\bibinfo {year} {2006})}\BibitemShut
  {NoStop}%
\bibitem [{\citenamefont {Comunas}\ \emph {et~al.}(2013)\citenamefont
  {Comunas}, \citenamefont {Paredes}, \citenamefont {Gaciño}, \citenamefont
  {Fernández}, \citenamefont {Bazile}, \citenamefont {Boned}, \citenamefont
  {Daridon}, \citenamefont {Galliero}, \citenamefont {Pauly}, \citenamefont
  {Harris}, \citenamefont {Assael},\ and\ \citenamefont {Mylona}}]{com13}%
  \BibitemOpen
  \bibfield  {author} {\bibinfo {author} {\bibfnamefont {M.~J.~P.}\
  \bibnamefont {Comunas}}, \bibinfo {author} {\bibfnamefont {X.}~\bibnamefont
  {Paredes}}, \bibinfo {author} {\bibfnamefont {F.~M.}\ \bibnamefont {Gaciño}},
  \bibinfo {author} {\bibfnamefont {J.}~\bibnamefont {Fernández}}, \bibinfo
  {author} {\bibfnamefont {J.~P.}\ \bibnamefont {Bazile}}, \bibinfo {author}
  {\bibfnamefont {C.}~\bibnamefont {Boned}}, \bibinfo {author} {\bibfnamefont
  {J.~L.}\ \bibnamefont {Daridon}}, \bibinfo {author} {\bibfnamefont
  {G.}~\bibnamefont {Galliero}}, \bibinfo {author} {\bibfnamefont
  {J.}~\bibnamefont {Pauly}}, \bibinfo {author} {\bibfnamefont {K.~R.}\
  \bibnamefont {Harris}}, \bibinfo {author} {\bibfnamefont {M.~J.}\
  \bibnamefont {Assael}}, \ and\ \bibinfo {author} {\bibfnamefont {S.~K.}\
  \bibnamefont {Mylona}},\ }\bibfield  {title} {\enquote {\bibinfo {title}
  {Reference correlation of the viscosity of squalane from 273 to 373 {K} at
  0.1 {MPa}},}\ }\href {\doibase http://dx.doi.org/10.1063/1.4812573}
  {\bibfield  {journal} {\bibinfo  {journal} {J. Phys. Chem. Ref. Data}\
  }\textbf {\bibinfo {volume} {42}},\ \bibinfo {pages} {033101} (\bibinfo
  {year} {2013})}\BibitemShut {NoStop}%
\bibitem [{\citenamefont {Spikes}\ and\ \citenamefont {Jie}(2014)}]{spi14}%
  \BibitemOpen
  \bibfield  {author} {\bibinfo {author} {\bibfnamefont {Hugh}\ \bibnamefont
  {Spikes}}\ and\ \bibinfo {author} {\bibfnamefont {Zhang}\ \bibnamefont
  {Jie}},\ }\bibfield  {title} {\enquote {\bibinfo {title} {History, origins
  and prediction of elastohydrodynamic friction},}\ }\href {\doibase
  10.1007/s11249-014-0396-y} {\bibfield  {journal} {\bibinfo  {journal}
  {Tribol. Lett.}\ }\textbf {\bibinfo {volume} {56}},\ \bibinfo {pages} {1--25}
  (\bibinfo {year} {2014})}\BibitemShut {NoStop}%
\bibitem [{\citenamefont {Schmidt}\ \emph {et~al.}(2015)\citenamefont
  {Schmidt}, \citenamefont {Pagnutti}, \citenamefont {Curran}, \citenamefont
  {Singh}, \citenamefont {Trusler}, \citenamefont {Maitland},\ and\
  \citenamefont {McBride-Wright}}]{sch15a}%
  \BibitemOpen
  \bibfield  {author} {\bibinfo {author} {\bibfnamefont {Kurt A.~G.}\
  \bibnamefont {Schmidt}}, \bibinfo {author} {\bibfnamefont {Doug}\
  \bibnamefont {Pagnutti}}, \bibinfo {author} {\bibfnamefont {Meghan~D.}\
  \bibnamefont {Curran}}, \bibinfo {author} {\bibfnamefont {Anil}\ \bibnamefont
  {Singh}}, \bibinfo {author} {\bibfnamefont {J.~P.~Martin}\ \bibnamefont
  {Trusler}}, \bibinfo {author} {\bibfnamefont {Geoffrey~C.}\ \bibnamefont
  {Maitland}}, \ and\ \bibinfo {author} {\bibfnamefont {Mark}\ \bibnamefont
  {McBride-Wright}},\ }\bibfield  {title} {\enquote {\bibinfo {title} {New
  experimental data and reference models for the viscosity and density of
  squalane},}\ }\href@noop {} {\bibfield  {journal} {\bibinfo  {journal} {J.
  Chem. Eng. Data}\ }\textbf {\bibinfo {volume} {60}},\ \bibinfo {pages}
  {137--150} (\bibinfo {year} {2015})}\BibitemShut {NoStop}%
\bibitem [{\citenamefont {Bair}\ \emph {et~al.}(2002)\citenamefont {Bair},
  \citenamefont {McCabe},\ and\ \citenamefont {Cummings}}]{bai02}%
  \BibitemOpen
  \bibfield  {author} {\bibinfo {author} {\bibfnamefont {Scott}\ \bibnamefont
  {Bair}}, \bibinfo {author} {\bibfnamefont {Clare}\ \bibnamefont {McCabe}}, \
  and\ \bibinfo {author} {\bibfnamefont {Peter~T.}\ \bibnamefont {Cummings}},\
  }\bibfield  {title} {\enquote {\bibinfo {title} {Comparison of nonequilibrium
  molecular dynamics with experimental measurements in the nonlinear
  shear-thinning regime},}\ }\href {\doibase 10.1103/PhysRevLett.88.058302}
  {\bibfield  {journal} {\bibinfo  {journal} {Phys. Rev. Lett.}\ }\textbf
  {\bibinfo {volume} {88}},\ \bibinfo {pages} {058302} (\bibinfo {year}
  {2002})}\BibitemShut {NoStop}%
\bibitem [{\citenamefont {Wang}\ \emph {et~al.}(2005)\citenamefont {Wang},
  \citenamefont {Shahriari},\ and\ \citenamefont {Richert}}]{wan05}%
  \BibitemOpen
  \bibfield  {author} {\bibinfo {author} {\bibfnamefont {L.~M.}\ \bibnamefont
  {Wang}}, \bibinfo {author} {\bibfnamefont {S.}~\bibnamefont {Shahriari}}, \
  and\ \bibinfo {author} {\bibfnamefont {R.}~\bibnamefont {Richert}},\
  }\bibfield  {title} {\enquote {\bibinfo {title} {Diluent effects on the
  {Debye-type} dielectric relaxation in viscous monohydroxy alcohols},}\ }\href
  {\doibase 10.1021/jp054542k} {\bibfield  {journal} {\bibinfo  {journal} {J.
  Phys. Chem. B}\ }\textbf {\bibinfo {volume} {109}},\ \bibinfo {pages}
  {23255--23262} (\bibinfo {year} {2005})}\BibitemShut {NoStop}%
\bibitem [{\citenamefont {Brocklehurst}\ and\ \citenamefont
  {Young}(1999)}]{bro99}%
  \BibitemOpen
  \bibfield  {author} {\bibinfo {author} {\bibfnamefont {B.}~\bibnamefont
  {Brocklehurst}}\ and\ \bibinfo {author} {\bibfnamefont {R.N.}\ \bibnamefont
  {Young}},\ }\bibfield  {title} {\enquote {\bibinfo {title} {Rotation of
  aromatic hydrocarbons in viscous alkanes. 2. {Hindered} rotation in
  squalane},}\ }\href {\doibase 10.1021/jp9843095} {\bibfield  {journal}
  {\bibinfo  {journal} {J. Phys. Chem. A}\ }\textbf {\bibinfo {volume} {103}},\
  \bibinfo {pages} {3818--3824} (\bibinfo {year} {1999})}\BibitemShut {NoStop}%
\bibitem [{\citenamefont {Kowert}\ and\ \citenamefont {Watson}(2011)}]{kow11}%
  \BibitemOpen
  \bibfield  {author} {\bibinfo {author} {\bibfnamefont {Bruce~A.}\
  \bibnamefont {Kowert}}\ and\ \bibinfo {author} {\bibfnamefont {Michael~B.}\
  \bibnamefont {Watson}},\ }\bibfield  {title} {\enquote {\bibinfo {title}
  {Diffusion of organic solutes in squalane},}\ }\href@noop {} {\bibfield
  {journal} {\bibinfo  {journal} {J. Phys. Chem. B}\ }\textbf {\bibinfo
  {volume} {115}},\ \bibinfo {pages} {9687--9694} (\bibinfo {year}
  {2011})}\BibitemShut {NoStop}%
\bibitem [{\citenamefont {Saecker}\ \emph {et~al.}(1991)\citenamefont
  {Saecker}, \citenamefont {Govoni}, \citenamefont {Kowalski}, \citenamefont
  {King},\ and\ \citenamefont {Nathanson}}]{sae91}%
  \BibitemOpen
  \bibfield  {author} {\bibinfo {author} {\bibfnamefont {M.~E.}\ \bibnamefont
  {Saecker}}, \bibinfo {author} {\bibfnamefont {S.~T.}\ \bibnamefont {Govoni}},
  \bibinfo {author} {\bibfnamefont {D.~V.}\ \bibnamefont {Kowalski}}, \bibinfo
  {author} {\bibfnamefont {M.~E.}\ \bibnamefont {King}}, \ and\ \bibinfo
  {author} {\bibfnamefont {G.~M.}\ \bibnamefont {Nathanson}},\ }\bibfield
  {title} {\enquote {\bibinfo {title} {Molecular beam scattering from liquid
  surfaces},}\ }\href@noop {} {\bibfield  {journal} {\bibinfo  {journal}
  {Science}\ }\textbf {\bibinfo {volume} {252}},\ \bibinfo {pages} {1421--1524}
  (\bibinfo {year} {1991})}\BibitemShut {NoStop}%
\bibitem [{\citenamefont {K{\"o}hler}\ \emph {et~al.}(2006)\citenamefont
  {K{\"o}hler}, \citenamefont {Reed}, \citenamefont {Westacott},\ and\
  \citenamefont {McKendrick}}]{koh06}%
  \BibitemOpen
  \bibfield  {author} {\bibinfo {author} {\bibfnamefont {S.~P.~K.}\
  \bibnamefont {K{\"o}hler}}, \bibinfo {author} {\bibfnamefont {S.~K.}\
  \bibnamefont {Reed}}, \bibinfo {author} {\bibfnamefont {R.~E.}\ \bibnamefont
  {Westacott}}, \ and\ \bibinfo {author} {\bibfnamefont {K.~G.}\ \bibnamefont
  {McKendrick}},\ }\bibfield  {title} {\enquote {\bibinfo {title} {Molecular
  dynamics study to identify the reactive sites of a liquid squalane
  surface},}\ }\href@noop {} {\bibfield  {journal} {\bibinfo  {journal} {J.
  Phys. Chem. B}\ }\textbf {\bibinfo {volume} {110}},\ \bibinfo {pages}
  {11717--11724} (\bibinfo {year} {2006})}\BibitemShut {NoStop}%
\bibitem [{\citenamefont {Barlow}\ and\ \citenamefont
  {Erginsav}(1972)}]{bar72}%
  \BibitemOpen
  \bibfield  {author} {\bibinfo {author} {\bibfnamefont {A.~J.}\ \bibnamefont
  {Barlow}}\ and\ \bibinfo {author} {\bibfnamefont {A.}~\bibnamefont
  {Erginsav}},\ }\bibfield  {title} {\enquote {\bibinfo {title} {Viscoelastic
  retardation of supercooled liquids},}\ }\href {\doibase
  10.1098/rspa.1972.0039} {\bibfield  {journal} {\bibinfo  {journal} {{Proc.
  Roy. Soc.}}\ }\textbf {\bibinfo {volume} {A327}},\ \bibinfo {pages}
  {175--190} (\bibinfo {year} {1972})}\BibitemShut {NoStop}%
\bibitem [{\citenamefont {Jakobsen}\ \emph {et~al.}(2011)\citenamefont
  {Jakobsen}, \citenamefont {Niss}, \citenamefont {Maggi}, \citenamefont
  {Olsen}, \citenamefont {Christensen},\ and\ \citenamefont {Dyre}}]{jak11}%
  \BibitemOpen
  \bibfield  {author} {\bibinfo {author} {\bibfnamefont {B.}~\bibnamefont
  {Jakobsen}}, \bibinfo {author} {\bibfnamefont {K.}~\bibnamefont {Niss}},
  \bibinfo {author} {\bibfnamefont {C.}~\bibnamefont {Maggi}}, \bibinfo
  {author} {\bibfnamefont {N.~B.}\ \bibnamefont {Olsen}}, \bibinfo {author}
  {\bibfnamefont {T.}~\bibnamefont {Christensen}}, \ and\ \bibinfo {author}
  {\bibfnamefont {J.~C.}\ \bibnamefont {Dyre}},\ }\bibfield  {title} {\enquote
  {\bibinfo {title} {Beta relaxation in the shear mechanics of viscous liquids:
  {Phenomenology} and network modeling of the alpha-beta merging region},}\
  }\href@noop {} {\bibfield  {journal} {\bibinfo  {journal} {J. Non-Cryst.
  Solids}\ }\textbf {\bibinfo {volume} {357}},\ \bibinfo {pages} {267--273}
  (\bibinfo {year} {2011})}\BibitemShut {NoStop}%
\bibitem [{\citenamefont {Christensen}\ and\ \citenamefont
  {Olsen}(1995)}]{chr95}%
  \BibitemOpen
  \bibfield  {author} {\bibinfo {author} {\bibfnamefont {T.}~\bibnamefont
  {Christensen}}\ and\ \bibinfo {author} {\bibfnamefont {N.~B.}\ \bibnamefont
  {Olsen}},\ }\bibfield  {title} {\enquote {\bibinfo {title} {A rheometer for
  the measurement of a high shear modulus covering more than seven decades of
  frequency below 50 {kHz }},}\ }\href@noop {} {\bibfield  {journal} {\bibinfo
  {journal} {Rev. Sci. Instrum.}\ }\textbf {\bibinfo {volume} {66}},\ \bibinfo
  {pages} {5019--5031} (\bibinfo {year} {1995})}\BibitemShut {NoStop}%
\bibitem [{\citenamefont {Hecksher}\ \emph {et~al.}(2013)\citenamefont
  {Hecksher}, \citenamefont {Olsen}, \citenamefont {Nelson}, \citenamefont
  {Dyre},\ and\ \citenamefont {Christensen}}]{hec13}%
  \BibitemOpen
  \bibfield  {author} {\bibinfo {author} {\bibfnamefont {T.}~\bibnamefont
  {Hecksher}}, \bibinfo {author} {\bibfnamefont {N.~B.}\ \bibnamefont {Olsen}},
  \bibinfo {author} {\bibfnamefont {K.~A.}\ \bibnamefont {Nelson}}, \bibinfo
  {author} {\bibfnamefont {J.~C.}\ \bibnamefont {Dyre}}, \ and\ \bibinfo
  {author} {\bibfnamefont {T.}~\bibnamefont {Christensen}},\ }\bibfield
  {title} {\enquote {\bibinfo {title} {Mechanical spectra of glass-forming
  liquids. i. low-frequency bulk and shear moduli of {DC704} and {5-PPE}
  measured by piezoceramic transducers},}\ }\href@noop {} {\bibfield  {journal}
  {\bibinfo  {journal} {J. Chem. Phys.}\ }\textbf {\bibinfo {volume} {138}},\
  \bibinfo {pages} {12A543} (\bibinfo {year} {2013})}\BibitemShut {NoStop}%
\bibitem [{\citenamefont {Igarashi}\ \emph
  {et~al.}(2008{\natexlab{a}})\citenamefont {Igarashi}, \citenamefont
  {Christensen}, \citenamefont {Larsen}, \citenamefont {Olsen}, \citenamefont
  {Pedersen}, \citenamefont {Rasmussen},\ and\ \citenamefont {Dyre}}]{iga08a}%
  \BibitemOpen
  \bibfield  {author} {\bibinfo {author} {\bibfnamefont {Brian}\ \bibnamefont
  {Igarashi}}, \bibinfo {author} {\bibfnamefont {Tage}\ \bibnamefont
  {Christensen}}, \bibinfo {author} {\bibfnamefont {Ebbe~H.}\ \bibnamefont
  {Larsen}}, \bibinfo {author} {\bibfnamefont {Niels~Boye}\ \bibnamefont
  {Olsen}}, \bibinfo {author} {\bibfnamefont {Ib~H.}\ \bibnamefont {Pedersen}},
  \bibinfo {author} {\bibfnamefont {Torben}\ \bibnamefont {Rasmussen}}, \ and\
  \bibinfo {author} {\bibfnamefont {Jeppe~C.}\ \bibnamefont {Dyre}},\
  }\bibfield  {title} {\enquote {\bibinfo {title} {A cryostat and temperature
  control system optimized for measuring relaxations of glass-forming
  liquids},}\ }\href {\doibase http://dx.doi.org/10.1063/1.2903419} {\bibfield
  {journal} {\bibinfo  {journal} {Rev. Sci. Instrum.}\ }\textbf {\bibinfo
  {volume} {79}},\ \bibinfo {eid} {045105} (\bibinfo {year}
  {2008}{\natexlab{a}})}\BibitemShut {NoStop}%
\bibitem [{\citenamefont {Igarashi}\ \emph
  {et~al.}(2008{\natexlab{b}})\citenamefont {Igarashi}, \citenamefont
  {Christensen}, \citenamefont {Larsen}, \citenamefont {Olsen}, \citenamefont
  {Pedersen}, \citenamefont {Rasmussen},\ and\ \citenamefont {Dyre}}]{iga08b}%
  \BibitemOpen
  \bibfield  {author} {\bibinfo {author} {\bibfnamefont {Brian}\ \bibnamefont
  {Igarashi}}, \bibinfo {author} {\bibfnamefont {Tage}\ \bibnamefont
  {Christensen}}, \bibinfo {author} {\bibfnamefont {Ebbe~H.}\ \bibnamefont
  {Larsen}}, \bibinfo {author} {\bibfnamefont {Niels~Boye}\ \bibnamefont
  {Olsen}}, \bibinfo {author} {\bibfnamefont {Ib~H.}\ \bibnamefont {Pedersen}},
  \bibinfo {author} {\bibfnamefont {Torben}\ \bibnamefont {Rasmussen}}, \ and\
  \bibinfo {author} {\bibfnamefont {Jeppe~C.}\ \bibnamefont {Dyre}},\
  }\bibfield  {title} {\enquote {\bibinfo {title} {An impedance-measurement
  setup optimized for measuring relaxations of glass-forming liquids},}\ }\href
  {\doibase http://dx.doi.org/10.1063/1.2906401} {\bibfield  {journal}
  {\bibinfo  {journal} {Rev. Sci. Instrum.}\ }\textbf {\bibinfo {volume}
  {79}},\ \bibinfo {eid} {045106} (\bibinfo {year}
  {2008}{\natexlab{b}})}\BibitemShut {NoStop}%
\bibitem [{dat()}]{datarepository}%
  \BibitemOpen
  \href@noop {} {}\bibinfo {note} {Details about the data can be obtained from
  the {\it Glass and Time} Data repository, found online at
  http://glass.ruc.dk/data.}\BibitemShut {Stop}%
\bibitem [{\citenamefont {Saglanmak}\ \emph {et~al.}(2010)\citenamefont
  {Saglanmak}, \citenamefont {Nielsen}, \citenamefont {Olsen}, \citenamefont
  {Dyre},\ and\ \citenamefont {Nissa}}]{sag10}%
  \BibitemOpen
  \bibfield  {author} {\bibinfo {author} {\bibfnamefont {N.}~\bibnamefont
  {Saglanmak}}, \bibinfo {author} {\bibfnamefont {A.~I.}\ \bibnamefont
  {Nielsen}}, \bibinfo {author} {\bibfnamefont {N.~B.}\ \bibnamefont {Olsen}},
  \bibinfo {author} {\bibfnamefont {J.~C.}\ \bibnamefont {Dyre}}, \ and\
  \bibinfo {author} {\bibfnamefont {K.}~\bibnamefont {Nissa}},\ }\bibfield
  {title} {\enquote {\bibinfo {title} {An electrical circuit model of the
  alpha-beta merging seen in dielectric relaxation of ultraviscous liquids},}\
  }\href@noop {} {\bibfield  {journal} {\bibinfo  {journal} {J. Chem. Phys.}\
  }\textbf {\bibinfo {volume} {132}},\ \bibinfo {pages} {024503} (\bibinfo
  {year} {2010})}\BibitemShut {NoStop}%
\bibitem [{\citenamefont {Jakobsen}\ \emph {et~al.}(2012)\citenamefont
  {Jakobsen}, \citenamefont {Hecksher}, \citenamefont {Christensen},
  \citenamefont {Olsen}, \citenamefont {Dyre},\ and\ \citenamefont
  {Niss}}]{jak12}%
  \BibitemOpen
  \bibfield  {author} {\bibinfo {author} {\bibfnamefont {B.}~\bibnamefont
  {Jakobsen}}, \bibinfo {author} {\bibfnamefont {T.}~\bibnamefont {Hecksher}},
  \bibinfo {author} {\bibfnamefont {T.}~\bibnamefont {Christensen}}, \bibinfo
  {author} {\bibfnamefont {N.~B.}\ \bibnamefont {Olsen}}, \bibinfo {author}
  {\bibfnamefont {J.~C.}\ \bibnamefont {Dyre}}, \ and\ \bibinfo {author}
  {\bibfnamefont {K.}~\bibnamefont {Niss}},\ }\bibfield  {title} {\enquote
  {\bibinfo {title} {Identical temperature dependence of the time scales of
  several linear-response functions of two glass-forming liquids},}\
  }\href@noop {} {\bibfield  {journal} {\bibinfo  {journal} {J. Chem. Phys.}\
  }\textbf {\bibinfo {volume} {136}},\ \bibinfo {pages} {081102} (\bibinfo
  {year} {2012})}\BibitemShut {NoStop}%
\bibitem [{\citenamefont {Christiansen}(1978)}]{pvc78}%
  \BibitemOpen
  \bibfield  {author} {\bibinfo {author} {\bibfnamefont {P.~V.}\ \bibnamefont
  {Christiansen}},\ }\href@noop {} {\emph {\bibinfo {title} {Dynamik og
  Diagrammer}}}\ (\bibinfo  {publisher} {IMFUFA tekst No. 8, Roskilde
  University},\ \bibinfo {address} {Roskilde},\ \bibinfo {year}
  {1978})\BibitemShut {NoStop}%
\bibitem [{\citenamefont {Karnopp}\ \emph {et~al.}(2006)\citenamefont
  {Karnopp}, \citenamefont {Margolis},\ and\ \citenamefont
  {Rosenberg}}]{systemdyn}%
  \BibitemOpen
  \bibfield  {author} {\bibinfo {author} {\bibfnamefont {D.~C.}\ \bibnamefont
  {Karnopp}}, \bibinfo {author} {\bibfnamefont {D.~L.}\ \bibnamefont
  {Margolis}}, \ and\ \bibinfo {author} {\bibfnamefont {R.~C.}\ \bibnamefont
  {Rosenberg}},\ }\href@noop {} {\emph {\bibinfo {title} {System Dynamics --
  Modeling and Simulation of Mechatronic Systems}}},\ \bibinfo {edition} {4th}\
  ed.\ (\bibinfo  {publisher} {Wiley},\ \bibinfo {address} {New York},\
  \bibinfo {year} {2006})\BibitemShut {NoStop}%
\bibitem [{\citenamefont {Paynter}(1961)}]{paynter}%
  \BibitemOpen
  \bibfield  {author} {\bibinfo {author} {\bibfnamefont {H.}~\bibnamefont
  {Paynter}},\ }\href@noop {} {\emph {\bibinfo {title} {Analysis and Design of
  Engineering Systems}}}\ (\bibinfo  {publisher} {MIT Press, Cambridge,
  Massachusetts},\ \bibinfo {year} {1961})\BibitemShut {NoStop}%
\bibitem [{\citenamefont {Oster}\ \emph {et~al.}(1971)\citenamefont {Oster},
  \citenamefont {Perelson},\ and\ \citenamefont {Katchalsky}}]{ost71}%
  \BibitemOpen
  \bibfield  {author} {\bibinfo {author} {\bibfnamefont {G.}~\bibnamefont
  {Oster}}, \bibinfo {author} {\bibfnamefont {A.}~\bibnamefont {Perelson}}, \
  and\ \bibinfo {author} {\bibfnamefont {A.}~\bibnamefont {Katchalsky}},\
  }\bibfield  {title} {\enquote {\bibinfo {title} {Network thermodynamics},}\
  }\href@noop {} {\bibfield  {journal} {\bibinfo  {journal} {Nature}\ }\textbf
  {\bibinfo {volume} {234}},\ \bibinfo {pages} {393--399} (\bibinfo {year}
  {1971})}\BibitemShut {NoStop}%
\bibitem [{\citenamefont {Schr{\o}der}\ \emph {et~al.}(2009)\citenamefont
  {Schr{\o}der}, \citenamefont {Bailey}, \citenamefont {Pedersen},
  \citenamefont {Gnan},\ and\ \citenamefont {Dyre}}]{III}%
  \BibitemOpen
  \bibfield  {author} {\bibinfo {author} {\bibfnamefont {T.~B.}\ \bibnamefont
  {Schr{\o}der}}, \bibinfo {author} {\bibfnamefont {N.~P.}\ \bibnamefont
  {Bailey}}, \bibinfo {author} {\bibfnamefont {U.~R.}\ \bibnamefont
  {Pedersen}}, \bibinfo {author} {\bibfnamefont {N.}~\bibnamefont {Gnan}}, \
  and\ \bibinfo {author} {\bibfnamefont {J.~C.}\ \bibnamefont {Dyre}},\
  }\bibfield  {title} {\enquote {\bibinfo {title} {{Pressure-Energy
  Correlations in Liquids. III. Statistical Mechanics and Thermodynamics of
  Liquids with Hidden Scale Invariance}},}\ }\href@noop {} {\bibfield
  {journal} {\bibinfo  {journal} {J. Chem. Phys.}\ }\textbf {\bibinfo {volume}
  {131}},\ \bibinfo {pages} {234503} (\bibinfo {year} {2009})}\BibitemShut
  {NoStop}%
\bibitem [{\citenamefont {Ferry}(1970)}]{ferry}%
  \BibitemOpen
  \bibfield  {author} {\bibinfo {author} {\bibfnamefont {J.~D.}\ \bibnamefont
  {Ferry}},\ }\href@noop {} {\emph {\bibinfo {title} {Viscoelastic Properties
  of Polymers, 2nd Edition}}}\ (\bibinfo  {publisher} {Wiley, New York},\
  \bibinfo {year} {1970})\BibitemShut {NoStop}%
\bibitem [{\citenamefont {Richert}(2015)}]{ric15}%
  \BibitemOpen
  \bibfield  {author} {\bibinfo {author} {\bibfnamefont {R.}~\bibnamefont
  {Richert}},\ }\bibfield  {title} {\enquote {\bibinfo {title} {Supercooled
  liquids and glasses by dielectric relaxation spectroscopy},}\ }\href@noop {}
  {\bibfield  {journal} {\bibinfo  {journal} {Adv. Chem. Phys.}\ }\textbf
  {\bibinfo {volume} {156}},\ \bibinfo {pages} {101--195} (\bibinfo {year}
  {2015})}\BibitemShut {NoStop}%
\bibitem [{\citenamefont {DiMarzio}\ and\ \citenamefont
  {Bishop}(1974)}]{dim74}%
  \BibitemOpen
  \bibfield  {author} {\bibinfo {author} {\bibfnamefont {E.~A.}\ \bibnamefont
  {DiMarzio}}\ and\ \bibinfo {author} {\bibfnamefont {M.}~\bibnamefont
  {Bishop}},\ }\bibfield  {title} {\enquote {\bibinfo {title} {Connection
  between the macroscopic electric and mechanical susceptibilities},}\ }\href
  {\doibase http://dx.doi.org/10.1063/1.1680822} {\bibfield  {journal}
  {\bibinfo  {journal} {J. Chem. Phys.}\ }\textbf {\bibinfo {volume} {60}},\
  \bibinfo {pages} {3802--3811} (\bibinfo {year} {1974})}\BibitemShut {NoStop}%
\bibitem [{\citenamefont {Niss}\ \emph {et~al.}(2005)\citenamefont {Niss},
  \citenamefont {Jakobsen},\ and\ \citenamefont {Olsen}}]{nis05}%
  \BibitemOpen
  \bibfield  {author} {\bibinfo {author} {\bibfnamefont {K.}~\bibnamefont
  {Niss}}, \bibinfo {author} {\bibfnamefont {B.}~\bibnamefont {Jakobsen}}, \
  and\ \bibinfo {author} {\bibfnamefont {N.~B.}\ \bibnamefont {Olsen}},\
  }\bibfield  {title} {\enquote {\bibinfo {title} {Dielectric and shear
  mechanical relaxations in glass-forming liquids: {A test of the
  Gemant-DiMarzio-Bishop} model},}\ }\href@noop {} {\bibfield  {journal}
  {\bibinfo  {journal} {J. Chem. Phys.}\ }\textbf {\bibinfo {volume} {123}},\
  \bibinfo {pages} {234510} (\bibinfo {year} {2005})}\BibitemShut {NoStop}%
\bibitem [{\citenamefont {Buchenau}(2011)}]{buc11}%
  \BibitemOpen
  \bibfield  {author} {\bibinfo {author} {\bibfnamefont {U.}~\bibnamefont
  {Buchenau}},\ }\bibfield  {title} {\enquote {\bibinfo {title} {Shear and
  dielectric spectra in highly viscous liquids},}\ }\href {\doibase
  10.1103/PhysRevB.83.052201} {\bibfield  {journal} {\bibinfo  {journal} {Phys.
  Rev. B}\ }\textbf {\bibinfo {volume} {83}},\ \bibinfo {pages} {052201}
  (\bibinfo {year} {2011})}\BibitemShut {NoStop}%
\bibitem [{\citenamefont {Garcia-Bernabe}\ \emph {et~al.}(2015)\citenamefont
  {Garcia-Bernabe}, \citenamefont {Lidon-Roger}, \citenamefont {Sanchis},
  \citenamefont {Diaz-Calleja},\ and\ \citenamefont {del Castillo}}]{gar15a}%
  \BibitemOpen
  \bibfield  {author} {\bibinfo {author} {\bibfnamefont {A.}~\bibnamefont
  {Garcia-Bernabe}}, \bibinfo {author} {\bibfnamefont {J.~V.}\ \bibnamefont
  {Lidon-Roger}}, \bibinfo {author} {\bibfnamefont {M.~J.}\ \bibnamefont
  {Sanchis}}, \bibinfo {author} {\bibfnamefont {R.}~\bibnamefont
  {Diaz-Calleja}}, \ and\ \bibinfo {author} {\bibfnamefont {L.~F.}\
  \bibnamefont {del Castillo}},\ }\bibfield  {title} {\enquote {\bibinfo
  {title} {Interconversion algorithm between mechanical and dielectric
  relaxation measurements for acetate of \textit{cis} - and \textit{trans}
  -2-phenyl-5-hydroxymethyl-1,3-dioxane},}\ }\href {\doibase
  10.1103/PhysRevE.92.042307} {\bibfield  {journal} {\bibinfo  {journal} {Phys.
  Rev. E}\ }\textbf {\bibinfo {volume} {92}},\ \bibinfo {pages} {042307}
  (\bibinfo {year} {2015})}\BibitemShut {NoStop}%
\bibitem [{\citenamefont {Meixner}\ and\ \citenamefont {Reik}(1959)}]{meixner}%
  \BibitemOpen
  \bibfield  {author} {\bibinfo {author} {\bibfnamefont {J.}~\bibnamefont
  {Meixner}}\ and\ \bibinfo {author} {\bibfnamefont {H.G.}\ \bibnamefont
  {Reik}},\ }\bibfield  {title} {\enquote {\bibinfo {title} {Thermodynamik der
  irreversiblen prozesse},}\ }in\ \href {\doibase 10.1007/978-3-642-45912-2_4}
  {\emph {\bibinfo {booktitle} {Prinzipien der Thermodynamik und Statistik /
  Principles of Thermodynamics and Statistics}}},\ \bibinfo {series} {Handbuch
  der Physik / Encyclopedia of Physics}, Vol.\ \bibinfo {volume} {2 / 3 / 2},\
  \bibinfo {editor} {edited by\ \bibinfo {editor} {\bibfnamefont
  {S.}~\bibnamefont {Fl{\"u}gge}}}\ (\bibinfo  {publisher} {Springer Berlin
  Heidelberg},\ \bibinfo {year} {1959})\ pp.\ \bibinfo {pages}
  {413--523}\BibitemShut {NoStop}%
\bibitem [{\citenamefont {Bernal}(1964)}]{ber64}%
  \BibitemOpen
  \bibfield  {author} {\bibinfo {author} {\bibfnamefont {J.~D.}\ \bibnamefont
  {Bernal}},\ }\bibfield  {title} {\enquote {\bibinfo {title} {The {Bakerian}
  lecture, 1962. {The} structure of liquids},}\ }\href@noop {} {\bibfield
  {journal} {\bibinfo  {journal} {Proc. R. Soc. London Ser. A}\ }\textbf
  {\bibinfo {volume} {280}},\ \bibinfo {pages} {299--322} (\bibinfo {year}
  {1964})}\BibitemShut {NoStop}%
\bibitem [{\citenamefont {Jonscher}(1996)}]{jonscher}%
  \BibitemOpen
  \bibfield  {author} {\bibinfo {author} {\bibfnamefont {A.~K.}\ \bibnamefont
  {Jonscher}},\ }\href@noop {} {\emph {\bibinfo {title} {Universal Relaxation
  Law}}}\ (\bibinfo  {publisher} {Chelsea Dielectric Press, London},\ \bibinfo
  {year} {1996})\BibitemShut {NoStop}%
\bibitem [{\citenamefont {Cole}\ and\ \citenamefont {Cole}(1941)}]{col41}%
  \BibitemOpen
  \bibfield  {author} {\bibinfo {author} {\bibfnamefont {Kenneth~S.}\
  \bibnamefont {Cole}}\ and\ \bibinfo {author} {\bibfnamefont {Robert~H.}\
  \bibnamefont {Cole}},\ }\bibfield  {title} {\enquote {\bibinfo {title}
  {Dispersion and absorption in dielectrics {I}. {Alternating} current
  characteristics},}\ }\href {\doibase http://dx.doi.org/10.1063/1.1750906}
  {\bibfield  {journal} {\bibinfo  {journal} {J. Chem. Phys.}\ }\textbf
  {\bibinfo {volume} {9}},\ \bibinfo {pages} {341--351} (\bibinfo {year}
  {1941})}\BibitemShut {NoStop}%
\bibitem [{\citenamefont {Narayanaswamy}(1971)}]{nar71}%
  \BibitemOpen
  \bibfield  {author} {\bibinfo {author} {\bibfnamefont {O.~S.}\ \bibnamefont
  {Narayanaswamy}},\ }\bibfield  {title} {\enquote {\bibinfo {title} {{A Model
  of Structural Relaxation in Glass}},}\ }\href@noop {} {\bibfield  {journal}
  {\bibinfo  {journal} {J. Am. Ceram. Soc.}\ }\textbf {\bibinfo {volume}
  {54}},\ \bibinfo {pages} {491--498} (\bibinfo {year} {1971})}\BibitemShut
  {NoStop}%
\bibitem [{\citenamefont {Scherer}(1986)}]{scherer}%
  \BibitemOpen
  \bibfield  {author} {\bibinfo {author} {\bibfnamefont {G.~W.}\ \bibnamefont
  {Scherer}},\ }\href@noop {} {\emph {\bibinfo {title} {{Relaxation in Glass
  and Composites}}}}\ (\bibinfo  {publisher} {Wiley, New York},\ \bibinfo
  {year} {1986})\BibitemShut {NoStop}%
\bibitem [{\citenamefont {Hecksher}\ \emph {et~al.}(2015)\citenamefont
  {Hecksher}, \citenamefont {Olsen},\ and\ \citenamefont {Dyre}}]{hec15}%
  \BibitemOpen
  \bibfield  {author} {\bibinfo {author} {\bibfnamefont {Tina}\ \bibnamefont
  {Hecksher}}, \bibinfo {author} {\bibfnamefont {Niels~Boye}\ \bibnamefont
  {Olsen}}, \ and\ \bibinfo {author} {\bibfnamefont {Jeppe~C.}\ \bibnamefont
  {Dyre}},\ }\bibfield  {title} {\enquote {\bibinfo {title} {Communication:
  Direct tests of single-parameter aging},}\ }\href {\doibase
  http://dx.doi.org/10.1063/1.4923000} {\bibfield  {journal} {\bibinfo
  {journal} {J. Chem. Phys.}\ }\textbf {\bibinfo {volume} {142}},\ \bibinfo
  {pages} {241103} (\bibinfo {year} {2015})}\BibitemShut {NoStop}%
\bibitem [{\citenamefont {Dyre}(2015)}]{dyr15}%
  \BibitemOpen
  \bibfield  {author} {\bibinfo {author} {\bibfnamefont {J.~C.}\ \bibnamefont
  {Dyre}},\ }\bibfield  {title} {\enquote {\bibinfo {title} {Narayanaswamys
  1971 aging theory and material time},}\ }\href@noop {} {\bibfield  {journal}
  {\bibinfo  {journal} {J. Chem. Phys.}\ }\textbf {\bibinfo {volume} {143}},\
  \bibinfo {pages} {114507} (\bibinfo {year} {2015})}\BibitemShut {NoStop}%
\bibitem [{\citenamefont {Lamb}(1978)}]{lam78}%
  \BibitemOpen
  \bibfield  {author} {\bibinfo {author} {\bibfnamefont {J.}~\bibnamefont
  {Lamb}},\ }\bibfield  {title} {\enquote {\bibinfo {title} {Viscoelasticity
  and lubrication: A review of liquid properties},}\ }\href@noop {} {\bibfield
  {journal} {\bibinfo  {journal} {J. Rheol.}\ }\textbf {\bibinfo {volume}
  {22}},\ \bibinfo {pages} {317--347} (\bibinfo {year} {1978})}\BibitemShut
  {NoStop}%
\bibitem [{\citenamefont {Olsen}\ \emph {et~al.}(2001)\citenamefont {Olsen},
  \citenamefont {Christensen},\ and\ \citenamefont {Dyre}}]{ols01}%
  \BibitemOpen
  \bibfield  {author} {\bibinfo {author} {\bibfnamefont {N.~B}\ \bibnamefont
  {Olsen}}, \bibinfo {author} {\bibfnamefont {T.}~\bibnamefont {Christensen}},
  \ and\ \bibinfo {author} {\bibfnamefont {J.~C.}\ \bibnamefont {Dyre}},\
  }\bibfield  {title} {\enquote {\bibinfo {title} {{Time-Temperature
  Superposition in Viscous Liquids}},}\ }\href@noop {} {\bibfield  {journal}
  {\bibinfo  {journal} {Phys. Rev. Lett.}\ }\textbf {\bibinfo {volume} {86}},\
  \bibinfo {pages} {1271--1274} (\bibinfo {year} {2001})}\BibitemShut {NoStop}%
\bibitem [{\citenamefont {Nielsen}\ \emph {et~al.}(2009)\citenamefont
  {Nielsen}, \citenamefont {Christensen}, \citenamefont {Jakobsen},
  \citenamefont {Niss}, \citenamefont {Olsen}, \citenamefont {Richert},\ and\
  \citenamefont {Dyre}}]{nie09}%
  \BibitemOpen
  \bibfield  {author} {\bibinfo {author} {\bibfnamefont {Albena~I.}\
  \bibnamefont {Nielsen}}, \bibinfo {author} {\bibfnamefont {Tage}\
  \bibnamefont {Christensen}}, \bibinfo {author} {\bibfnamefont
  {Bo}~\bibnamefont {Jakobsen}}, \bibinfo {author} {\bibfnamefont {Kristine}\
  \bibnamefont {Niss}}, \bibinfo {author} {\bibfnamefont {Niels~Boye}\
  \bibnamefont {Olsen}}, \bibinfo {author} {\bibfnamefont {Ranko}\ \bibnamefont
  {Richert}}, \ and\ \bibinfo {author} {\bibfnamefont {Jeppe~C.}\ \bibnamefont
  {Dyre}},\ }\bibfield  {title} {\enquote {\bibinfo {title} {Prevalence of
  approximate $\sqrt{t}$ relaxation for the dielectric Î± process in viscous
  organic liquids},}\ }\href {\doibase http://dx.doi.org/10.1063/1.3098911}
  {\bibfield  {journal} {\bibinfo  {journal} {J. Chem. Phys.}\ }\textbf
  {\bibinfo {volume} {130}},\ \bibinfo {eid} {154508} (\bibinfo {year}
  {2009})}\BibitemShut {NoStop}%
\bibitem [{\citenamefont {Dyre}(2005)}]{dyr05}%
  \BibitemOpen
  \bibfield  {author} {\bibinfo {author} {\bibfnamefont {Jeppe~C.}\
  \bibnamefont {Dyre}},\ }\bibfield  {title} {\enquote {\bibinfo {title}
  {Solidity of viscous liquids. {III}. $\ensuremath{\alpha}$ relaxation},}\
  }\href {\doibase 10.1103/PhysRevE.72.011501} {\bibfield  {journal} {\bibinfo
  {journal} {Phys. Rev. E}\ }\textbf {\bibinfo {volume} {72}},\ \bibinfo
  {pages} {011501} (\bibinfo {year} {2005})}\BibitemShut {NoStop}%
\bibitem [{\citenamefont {Dyre}(2006{\natexlab{b}})}]{dyr06a}%
  \BibitemOpen
  \bibfield  {author} {\bibinfo {author} {\bibfnamefont {Jeppe~C.}\
  \bibnamefont {Dyre}},\ }\bibfield  {title} {\enquote {\bibinfo {title}
  {Solidity of viscous liquids. {IV}. {Density} fluctuations},}\ }\href
  {\doibase 10.1103/PhysRevE.74.021502} {\bibfield  {journal} {\bibinfo
  {journal} {Phys. Rev. E}\ }\textbf {\bibinfo {volume} {74}},\ \bibinfo
  {pages} {021502} (\bibinfo {year} {2006}{\natexlab{b}})}\BibitemShut
  {NoStop}%
\bibitem [{\citenamefont {Dyre}(2007)}]{dyr07}%
  \BibitemOpen
  \bibfield  {author} {\bibinfo {author} {\bibfnamefont {Jeppe~C.}\
  \bibnamefont {Dyre}},\ }\bibfield  {title} {\enquote {\bibinfo {title}
  {Solidity of viscous liquids. {V}. {Long}-wavelength dominance of the
  dynamics},}\ }\href {\doibase 10.1103/PhysRevE.76.041508} {\bibfield
  {journal} {\bibinfo  {journal} {Phys. Rev. E}\ }\textbf {\bibinfo {volume}
  {76}},\ \bibinfo {pages} {041508} (\bibinfo {year} {2007})}\BibitemShut
  {NoStop}%
\bibitem [{\citenamefont {Ooshida}\ \emph {et~al.}(2016)\citenamefont
  {Ooshida}, \citenamefont {Goto}, \citenamefont {Matsumoto},\ and\
  \citenamefont {Otsuki}}]{oos16}%
  \BibitemOpen
  \bibfield  {author} {\bibinfo {author} {\bibfnamefont {Takeshi}\ \bibnamefont
  {Ooshida}}, \bibinfo {author} {\bibfnamefont {Susumu}\ \bibnamefont {Goto}},
  \bibinfo {author} {\bibfnamefont {Takeshi}\ \bibnamefont {Matsumoto}}, \ and\
  \bibinfo {author} {\bibfnamefont {Michio}\ \bibnamefont {Otsuki}},\
  }\bibfield  {title} {\enquote {\bibinfo {title} {Insights from single-file
  diffusion into cooperativity in higher dimensions},}\ }\href {\doibase
  10.1142/S1793048015400019} {\bibfield  {journal} {\bibinfo  {journal}
  {Biophys. Rev. Lett.}\ }\textbf {\bibinfo {volume} {11}},\ \bibinfo {pages}
  {9--38} (\bibinfo {year} {2016})}\BibitemShut {NoStop}%
\bibitem [{\citenamefont {Barlow}\ \emph {et~al.}(1967)\citenamefont {Barlow},
  \citenamefont {Erginsav},\ and\ \citenamefont {Lamb}}]{BEL}%
  \BibitemOpen
  \bibfield  {author} {\bibinfo {author} {\bibfnamefont {A.~J.}\ \bibnamefont
  {Barlow}}, \bibinfo {author} {\bibfnamefont {A.}~\bibnamefont {Erginsav}}, \
  and\ \bibinfo {author} {\bibfnamefont {J.}~\bibnamefont {Lamb}},\ }\bibfield
  {title} {\enquote {\bibinfo {title} {Viscoelastic relaxation of supercooled
  liquids},}\ }\href@noop {} {\bibfield  {journal} {\bibinfo  {journal} {Proc.
  R. Soc. London}\ }\textbf {\bibinfo {volume} {298}},\ \bibinfo {pages} {481}
  (\bibinfo {year} {1967})}\BibitemShut {NoStop}%
\bibitem [{\citenamefont {Barlow}(1972)}]{bar72a}%
  \BibitemOpen
  \bibfield  {author} {\bibinfo {author} {\bibfnamefont {A.~J.}\ \bibnamefont
  {Barlow}},\ }\bibfield  {title} {\enquote {\bibinfo {title} {Viscoelastic
  relaxation of supercooled liquids},}\ }\href@noop {} {\bibfield  {journal}
  {\bibinfo  {journal} {Adv. Mol. Relax. Proc.}\ }\textbf {\bibinfo {volume}
  {3}},\ \bibinfo {pages} {256--266} (\bibinfo {year} {1972})}\BibitemShut
  {NoStop}%
\bibitem [{\citenamefont {Olsen}(1998)}]{ols98a}%
  \BibitemOpen
  \bibfield  {author} {\bibinfo {author} {\bibfnamefont {N.~B.}\ \bibnamefont
  {Olsen}},\ }\bibfield  {title} {\enquote {\bibinfo {title} {Scaling of
  $\beta$-relaxation in the equilibrium liquid state of sorbitol},}\
  }\href@noop {} {\bibfield  {journal} {\bibinfo  {journal} {J. Non-Cryst.
  Solids}\ }\textbf {\bibinfo {volume} {235}},\ \bibinfo {pages} {399--405}
  (\bibinfo {year} {1998})}\BibitemShut {NoStop}%
\bibitem [{\citenamefont {Olsen}\ \emph {et~al.}(2000)\citenamefont {Olsen},
  \citenamefont {Christensen},\ and\ \citenamefont {Dyre}}]{ols00}%
  \BibitemOpen
  \bibfield  {author} {\bibinfo {author} {\bibfnamefont {Niels~Boye}\
  \bibnamefont {Olsen}}, \bibinfo {author} {\bibfnamefont {Tage}\ \bibnamefont
  {Christensen}}, \ and\ \bibinfo {author} {\bibfnamefont {Jeppe~C.}\
  \bibnamefont {Dyre}},\ }\bibfield  {title} {\enquote {\bibinfo {title}
  {\ensuremath{\beta} relaxation of nonpolymeric liquids close to the glass
  transition},}\ }\href {\doibase 10.1103/PhysRevE.62.4435} {\bibfield
  {journal} {\bibinfo  {journal} {Phys. Rev. E}\ }\textbf {\bibinfo {volume}
  {62}},\ \bibinfo {pages} {4435--4438} (\bibinfo {year} {2000})}\BibitemShut
  {NoStop}%
\bibitem [{\citenamefont {Dyre}\ and\ \citenamefont {Olsen}(2003)}]{dyr03}%
  \BibitemOpen
  \bibfield  {author} {\bibinfo {author} {\bibfnamefont {J.~C.}\ \bibnamefont
  {Dyre}}\ and\ \bibinfo {author} {\bibfnamefont {N.~B.}\ \bibnamefont
  {Olsen}},\ }\bibfield  {title} {\enquote {\bibinfo {title} {Minimal model for
  beta relaxation in viscous liquids},}\ }\href@noop {} {\bibfield  {journal}
  {\bibinfo  {journal} {Phys. Rev. Lett.}\ }\textbf {\bibinfo {volume} {91}},\
  \bibinfo {pages} {155703} (\bibinfo {year} {2003})}\BibitemShut {NoStop}%
\bibitem [{\citenamefont {Johari}(1982)}]{joh82}%
  \BibitemOpen
  \bibfield  {author} {\bibinfo {author} {\bibfnamefont {G.~P.}\ \bibnamefont
  {Johari}},\ }\bibfield  {title} {\enquote {\bibinfo {title} {{Effect of
  Annealing on the Secondary Relaxations in Glasses}},}\ }\href@noop {}
  {\bibfield  {journal} {\bibinfo  {journal} {J. Chem. Phys.}\ }\textbf
  {\bibinfo {volume} {77}},\ \bibinfo {pages} {4619--4626} (\bibinfo {year}
  {1982})}\BibitemShut {NoStop}%
\bibitem [{\citenamefont {Qiao}\ \emph {et~al.}(2014)\citenamefont {Qiao},
  \citenamefont {Casalini},\ and\ \citenamefont {Pelletier}}]{qia14a}%
  \BibitemOpen
  \bibfield  {author} {\bibinfo {author} {\bibfnamefont {J.}~\bibnamefont
  {Qiao}}, \bibinfo {author} {\bibfnamefont {R.}~\bibnamefont {Casalini}}, \
  and\ \bibinfo {author} {\bibfnamefont {J.-M.}\ \bibnamefont {Pelletier}},\
  }\bibfield  {title} {\enquote {\bibinfo {title} {Effect of physical aging on
  {Johari-Goldstein} relaxation in {La}-based bulk metallic glass},}\
  }\href@noop {} {\bibfield  {journal} {\bibinfo  {journal} {J. Chem. Phys.}\
  }\textbf {\bibinfo {volume} {141}},\ \bibinfo {pages} {104510} (\bibinfo
  {year} {2014})}\BibitemShut {NoStop}%
\bibitem [{\citenamefont {Christensen}\ and\ \citenamefont
  {Olsen}(1994)}]{chr94}%
  \BibitemOpen
  \bibfield  {author} {\bibinfo {author} {\bibfnamefont {T.}~\bibnamefont
  {Christensen}}\ and\ \bibinfo {author} {\bibfnamefont {N.~B.}\ \bibnamefont
  {Olsen}},\ }\bibfield  {title} {\enquote {\bibinfo {title} {Determination of
  the frequency-dependent bulk modulus of glycerol using a piezoelectric
  spherical shell},}\ }\href {\doibase 10.1103/PhysRevB.49.15396} {\bibfield
  {journal} {\bibinfo  {journal} {Phys. Rev. B}\ }\textbf {\bibinfo {volume}
  {49}},\ \bibinfo {pages} {15396--15399} (\bibinfo {year} {1994})}\BibitemShut
  {NoStop}%
\bibitem [{\citenamefont {Gundermann}\ \emph {et~al.}(2014)\citenamefont
  {Gundermann}, \citenamefont {Niss}, \citenamefont {Christensen},
  \citenamefont {Dyre},\ and\ \citenamefont {Hecksher}}]{gun14}%
  \BibitemOpen
  \bibfield  {author} {\bibinfo {author} {\bibfnamefont {D.}~\bibnamefont
  {Gundermann}}, \bibinfo {author} {\bibfnamefont {K.}~\bibnamefont {Niss}},
  \bibinfo {author} {\bibfnamefont {T.}~\bibnamefont {Christensen}}, \bibinfo
  {author} {\bibfnamefont {J.~C.}\ \bibnamefont {Dyre}}, \ and\ \bibinfo
  {author} {\bibfnamefont {T.}~\bibnamefont {Hecksher}},\ }\bibfield  {title}
  {\enquote {\bibinfo {title} {The dynamic bulk modulus of three glass-forming
  liquids},}\ }\href@noop {} {\bibfield  {journal} {\bibinfo  {journal} {J.
  Chem. Phys.}\ }\textbf {\bibinfo {volume} {140}},\ \bibinfo {pages} {244508}
  (\bibinfo {year} {2014})}\BibitemShut {NoStop}%
\bibitem [{\citenamefont {Buchenau}(2012)}]{buc12b}%
  \BibitemOpen
  \bibfield  {author} {\bibinfo {author} {\bibfnamefont {U.}~\bibnamefont
  {Buchenau}},\ }\bibfield  {title} {\enquote {\bibinfo {title} {Bulk and shear
  relaxation in glasses and highly viscous liquids},}\ }\href {\doibase
  http://dx.doi.org/10.1063/1.4726459} {\bibfield  {journal} {\bibinfo
  {journal} {J. Chem. Phys.}\ }\textbf {\bibinfo {volume} {136}},\ \bibinfo
  {pages} {224512} (\bibinfo {year} {2012})}\BibitemShut {NoStop}%
\bibitem [{\citenamefont {Ellegaard}\ \emph {et~al.}(2007)\citenamefont
  {Ellegaard}, \citenamefont {Christensen}, \citenamefont {Christiansen},
  \citenamefont {Olsen}, \citenamefont {Pedersen}, \citenamefont
  {Schr{\o}der},\ and\ \citenamefont {Dyre}}]{ell07}%
  \BibitemOpen
  \bibfield  {author} {\bibinfo {author} {\bibfnamefont {N.~L.}\ \bibnamefont
  {Ellegaard}}, \bibinfo {author} {\bibfnamefont {T.}~\bibnamefont
  {Christensen}}, \bibinfo {author} {\bibfnamefont {P.~V.}\ \bibnamefont
  {Christiansen}}, \bibinfo {author} {\bibfnamefont {N.~B.}\ \bibnamefont
  {Olsen}}, \bibinfo {author} {\bibfnamefont {U.~R}\ \bibnamefont {Pedersen}},
  \bibinfo {author} {\bibfnamefont {T.~B.}\ \bibnamefont {Schr{\o}der}}, \ and\
  \bibinfo {author} {\bibfnamefont {J.~C.}\ \bibnamefont {Dyre}},\ }\bibfield
  {title} {\enquote {\bibinfo {title} {{Single-Order-Parameter Description of
  Glass-Forming Liquids: A One-Frequency Test}},}\ }\href@noop {} {\bibfield
  {journal} {\bibinfo  {journal} {J. Chem. Phys.}\ }\textbf {\bibinfo {volume}
  {126}},\ \bibinfo {pages} {074502} (\bibinfo {year} {2007})}\BibitemShut
  {NoStop}%
\bibitem [{\citenamefont {Bailey}\ \emph {et~al.}(2008)\citenamefont {Bailey},
  \citenamefont {Christensen}, \citenamefont {Jakobsen}, \citenamefont {Niss},
  \citenamefont {Olsen}, \citenamefont {Pedersen}, \citenamefont
  {Schr{\o}der},\ and\ \citenamefont {Dyre}}]{bai08}%
  \BibitemOpen
  \bibfield  {author} {\bibinfo {author} {\bibfnamefont {N.~P.}\ \bibnamefont
  {Bailey}}, \bibinfo {author} {\bibfnamefont {T.}~\bibnamefont {Christensen}},
  \bibinfo {author} {\bibfnamefont {B.}~\bibnamefont {Jakobsen}}, \bibinfo
  {author} {\bibfnamefont {K.}~\bibnamefont {Niss}}, \bibinfo {author}
  {\bibfnamefont {N.~B.}\ \bibnamefont {Olsen}}, \bibinfo {author}
  {\bibfnamefont {U.~R.}\ \bibnamefont {Pedersen}}, \bibinfo {author}
  {\bibfnamefont {T.~B.}\ \bibnamefont {Schr{\o}der}}, \ and\ \bibinfo {author}
  {\bibfnamefont {J.~C.}\ \bibnamefont {Dyre}},\ }\bibfield  {title} {\enquote
  {\bibinfo {title} {Glass-forming liquids: {One} or more "order"
  parameters?}}\ }\href@noop {} {\bibfield  {journal} {\bibinfo  {journal} {J.
  Phys. Condens. Matter}\ }\textbf {\bibinfo {volume} {20}},\ \bibinfo {pages}
  {244113} (\bibinfo {year} {2008})}\BibitemShut {NoStop}%
\bibitem [{\citenamefont {Gundermann}\ \emph {et~al.}(2011)\citenamefont
  {Gundermann}, \citenamefont {Pedersen}, \citenamefont {Hecksher},
  \citenamefont {Bailey}, \citenamefont {Jakobsen}, \citenamefont
  {Christensen}, \citenamefont {Olsen}, \citenamefont {Schr{\o}der},
  \citenamefont {Fragiadakis}, \citenamefont {Casalini}, \citenamefont
  {Roland}, \citenamefont {Dyre},\ and\ \citenamefont {Niss}}]{gun11}%
  \BibitemOpen
  \bibfield  {author} {\bibinfo {author} {\bibfnamefont {D.}~\bibnamefont
  {Gundermann}}, \bibinfo {author} {\bibfnamefont {U.~R.}\ \bibnamefont
  {Pedersen}}, \bibinfo {author} {\bibfnamefont {T.}~\bibnamefont {Hecksher}},
  \bibinfo {author} {\bibfnamefont {N.~P.}\ \bibnamefont {Bailey}}, \bibinfo
  {author} {\bibfnamefont {B.}~\bibnamefont {Jakobsen}}, \bibinfo {author}
  {\bibfnamefont {T.}~\bibnamefont {Christensen}}, \bibinfo {author}
  {\bibfnamefont {N.~B.}\ \bibnamefont {Olsen}}, \bibinfo {author}
  {\bibfnamefont {T.~B.}\ \bibnamefont {Schr{\o}der}}, \bibinfo {author}
  {\bibfnamefont {D.}~\bibnamefont {Fragiadakis}}, \bibinfo {author}
  {\bibfnamefont {R.}~\bibnamefont {Casalini}}, \bibinfo {author}
  {\bibfnamefont {C.~M.}\ \bibnamefont {Roland}}, \bibinfo {author}
  {\bibfnamefont {J.~C.}\ \bibnamefont {Dyre}}, \ and\ \bibinfo {author}
  {\bibfnamefont {K.}~\bibnamefont {Niss}},\ }\bibfield  {title} {\enquote
  {\bibinfo {title} {{Predicting the Density-Scaling Exponent of a
  Glass--Forming Liquid from Prigogine-Defay Ratio Measurements}},}\
  }\href@noop {} {\bibfield  {journal} {\bibinfo  {journal} {Nature Phys.}\
  }\textbf {\bibinfo {volume} {7}},\ \bibinfo {pages} {816--821} (\bibinfo
  {year} {2011})}\BibitemShut {NoStop}%
\bibitem [{\citenamefont {Pedersen}\ \emph {et~al.}(2011)\citenamefont
  {Pedersen}, \citenamefont {Gnan}, \citenamefont {Bailey}, \citenamefont
  {Schr{\o}der},\ and\ \citenamefont {Dyre}}]{ped11}%
  \BibitemOpen
  \bibfield  {author} {\bibinfo {author} {\bibfnamefont {U.~R.}\ \bibnamefont
  {Pedersen}}, \bibinfo {author} {\bibfnamefont {N.}~\bibnamefont {Gnan}},
  \bibinfo {author} {\bibfnamefont {N.~P.}\ \bibnamefont {Bailey}}, \bibinfo
  {author} {\bibfnamefont {T.~B.}\ \bibnamefont {Schr{\o}der}}, \ and\ \bibinfo
  {author} {\bibfnamefont {J.~C.}\ \bibnamefont {Dyre}},\ }\bibfield  {title}
  {\enquote {\bibinfo {title} {{Strongly Correlating Liquids and their
  Isomorphs}},}\ }\href@noop {} {\bibfield  {journal} {\bibinfo  {journal} {J.
  Non-Cryst. Solids}\ }\textbf {\bibinfo {volume} {357}},\ \bibinfo {pages}
  {320--328} (\bibinfo {year} {2011})}\BibitemShut {NoStop}%
\bibitem [{\citenamefont {Ingebrigtsen}\ \emph {et~al.}(2012)\citenamefont
  {Ingebrigtsen}, \citenamefont {Schr\o{}der},\ and\ \citenamefont
  {Dyre}}]{ing12}%
  \BibitemOpen
  \bibfield  {author} {\bibinfo {author} {\bibfnamefont {T.~S.}\ \bibnamefont
  {Ingebrigtsen}}, \bibinfo {author} {\bibfnamefont {T.~B.}\ \bibnamefont
  {Schr\o{}der}}, \ and\ \bibinfo {author} {\bibfnamefont {J.~C.}\ \bibnamefont
  {Dyre}},\ }\bibfield  {title} {\enquote {\bibinfo {title} {What is a simple
  liquid?}}\ }\href@noop {} {\bibfield  {journal} {\bibinfo  {journal} {Phys.
  Rev. X}\ }\textbf {\bibinfo {volume} {2}},\ \bibinfo {pages} {011011}
  (\bibinfo {year} {2012})}\BibitemShut {NoStop}%
\bibitem [{\citenamefont {Dyre}(2014)}]{dyr14}%
  \BibitemOpen
  \bibfield  {author} {\bibinfo {author} {\bibfnamefont {J.~C.}\ \bibnamefont
  {Dyre}},\ }\bibfield  {title} {\enquote {\bibinfo {title} {{Hidden Scale
  Invariance in Condensed Matter}},}\ }\href@noop {} {\bibfield  {journal}
  {\bibinfo  {journal} {J. Phys. Chem. B}\ }\textbf {\bibinfo {volume} {118}},\
  \bibinfo {pages} {10007--10024} (\bibinfo {year} {2014})}\BibitemShut
  {NoStop}%
\bibitem [{\citenamefont {Dyre}(2016)}]{dyr16}%
  \BibitemOpen
  \bibfield  {author} {\bibinfo {author} {\bibfnamefont {J.~C.}\ \bibnamefont
  {Dyre}},\ }\bibfield  {title} {\enquote {\bibinfo {title} {Simple {liquids}
  quasiuniversality and the hard-sphere paradigm},}\ }\href@noop {} {\bibfield
  {journal} {\bibinfo  {journal} {J. Phys.: Condens. Mat.}\ }\textbf {\bibinfo
  {volume} {28}},\ \bibinfo {pages} {323001} (\bibinfo {year}
  {2016})}\BibitemShut {NoStop}%
\bibitem [{\citenamefont {Niss}\ \emph {et~al.}(2012)\citenamefont {Niss},
  \citenamefont {Gundermann}, \citenamefont {Christensen},\ and\ \citenamefont
  {Dyre}}]{nis12}%
  \BibitemOpen
  \bibfield  {author} {\bibinfo {author} {\bibfnamefont {K.}~\bibnamefont
  {Niss}}, \bibinfo {author} {\bibfnamefont {D.}~\bibnamefont {Gundermann}},
  \bibinfo {author} {\bibfnamefont {T.}~\bibnamefont {Christensen}}, \ and\
  \bibinfo {author} {\bibfnamefont {J.~C.}\ \bibnamefont {Dyre}},\ }\bibfield
  {title} {\enquote {\bibinfo {title} {Dynamic thermal expansivity of liquids
  near the glass transition},}\ }\href@noop {} {\bibfield  {journal} {\bibinfo
  {journal} {Phys. Rev. E}\ }\textbf {\bibinfo {volume} {85}},\ \bibinfo
  {pages} {041501} (\bibinfo {year} {2012})}\BibitemShut {NoStop}%
\bibitem [{\citenamefont {Buchenau}(2016)}]{buc16}%
  \BibitemOpen
  \bibfield  {author} {\bibinfo {author} {\bibfnamefont {U.}~\bibnamefont
  {Buchenau}},\ }\bibfield  {title} {\enquote {\bibinfo {title} {Retardation
  and flow at the glass transition},}\ }\href {\doibase
  10.1103/PhysRevE.93.032608} {\bibfield  {journal} {\bibinfo  {journal} {Phys.
  Rev. E}\ }\textbf {\bibinfo {volume} {93}},\ \bibinfo {pages} {032608}
  (\bibinfo {year} {2016})}\BibitemShut {NoStop}%
\bibitem [{\citenamefont {Diezemann}\ \emph {et~al.}(1999)\citenamefont
  {Diezemann}, \citenamefont {Mohanty},\ and\ \citenamefont
  {Oppenheim}}]{die99}%
  \BibitemOpen
  \bibfield  {author} {\bibinfo {author} {\bibfnamefont {Gregor}\ \bibnamefont
  {Diezemann}}, \bibinfo {author} {\bibfnamefont {Udayan}\ \bibnamefont
  {Mohanty}}, \ and\ \bibinfo {author} {\bibfnamefont {Irwin}\ \bibnamefont
  {Oppenheim}},\ }\bibfield  {title} {\enquote {\bibinfo {title} {Slow
  secondary relaxation in a free-energy landscape model for relaxation in
  glass-forming liquids},}\ }\href {\doibase 10.1103/PhysRevE.59.2067}
  {\bibfield  {journal} {\bibinfo  {journal} {Phys. Rev. E}\ }\textbf {\bibinfo
  {volume} {59}},\ \bibinfo {pages} {2067--2083} (\bibinfo {year}
  {1999})}\BibitemShut {NoStop}%
\bibitem [{\citenamefont {B{\"o}hmer}\ \emph {et~al.}(2014)\citenamefont
  {B{\"o}hmer}, \citenamefont {Gainaru},\ and\ \citenamefont
  {Richert}}]{boh14a}%
  \BibitemOpen
  \bibfield  {author} {\bibinfo {author} {\bibfnamefont {R.}~\bibnamefont
  {B{\"o}hmer}}, \bibinfo {author} {\bibfnamefont {C.}~\bibnamefont {Gainaru}},
  \ and\ \bibinfo {author} {\bibfnamefont {R.}~\bibnamefont {Richert}},\
  }\bibfield  {title} {\enquote {\bibinfo {title} {Structure and dynamics of
  monohydroxy alcohols -- {Milestones} towards their microscopic
  understanding{, 100}\, years after {Debye}},}\ }\href {\doibase
  http://dx.doi.org/10.1016/j.physrep.2014.07.005} {\bibfield  {journal}
  {\bibinfo  {journal} {Phys. Rep.}\ }\textbf {\bibinfo {volume} {545}},\
  \bibinfo {pages} {125 -- 195} (\bibinfo {year} {2014})}\BibitemShut {NoStop}%
\bibitem [{\citenamefont {Gainaru}\ \emph {et~al.}(2014)\citenamefont
  {Gainaru}, \citenamefont {Figuli}, \citenamefont {Hecksher}, \citenamefont
  {Jakobsen}, \citenamefont {Dyre}, \citenamefont {Wilhelm},\ and\
  \citenamefont {B{\"o}hmer}}]{gai14}%
  \BibitemOpen
  \bibfield  {author} {\bibinfo {author} {\bibfnamefont {C.}~\bibnamefont
  {Gainaru}}, \bibinfo {author} {\bibfnamefont {R.}~\bibnamefont {Figuli}},
  \bibinfo {author} {\bibfnamefont {T.}~\bibnamefont {Hecksher}}, \bibinfo
  {author} {\bibfnamefont {B.}~\bibnamefont {Jakobsen}}, \bibinfo {author}
  {\bibfnamefont {J.~C.}\ \bibnamefont {Dyre}}, \bibinfo {author}
  {\bibfnamefont {M.}~\bibnamefont {Wilhelm}}, \ and\ \bibinfo {author}
  {\bibfnamefont {R.}~\bibnamefont {B{\"o}hmer}},\ }\bibfield  {title}
  {\enquote {\bibinfo {title} {Shear-modulus investigations of monohydroxy
  alcohols: {Evidence} for a short-chain-polymer rheological response},}\
  }\href {\doibase 10.1103/PhysRevLett.112.098301} {\bibfield  {journal}
  {\bibinfo  {journal} {Phys. Rev. Lett.}\ }\textbf {\bibinfo {volume} {112}},\
  \bibinfo {pages} {098301} (\bibinfo {year} {2014})}\BibitemShut {NoStop}%
\bibitem [{\citenamefont {Dyre}\ \emph {et~al.}(1996)\citenamefont {Dyre},
  \citenamefont {Olsen},\ and\ \citenamefont {Christensen}}]{dyr96}%
  \BibitemOpen
  \bibfield  {author} {\bibinfo {author} {\bibfnamefont {J.~C.}\ \bibnamefont
  {Dyre}}, \bibinfo {author} {\bibfnamefont {N.~B.}\ \bibnamefont {Olsen}}, \
  and\ \bibinfo {author} {\bibfnamefont {T.}~\bibnamefont {Christensen}},\
  }\bibfield  {title} {\enquote {\bibinfo {title} {Local elastic expansion
  model for viscous-flow activation energies of glass-forming molecular
  liquids},}\ }\href@noop {} {\bibfield  {journal} {\bibinfo  {journal} {Phys.
  Rev. B}\ }\textbf {\bibinfo {volume} {53}},\ \bibinfo {pages} {2171--2174}
  (\bibinfo {year} {1996})}\BibitemShut {NoStop}%
\bibitem [{\citenamefont {Hecksher}\ and\ \citenamefont {Dyre}(2015)}]{hec15a}%
  \BibitemOpen
  \bibfield  {author} {\bibinfo {author} {\bibfnamefont {T.}~\bibnamefont
  {Hecksher}}\ and\ \bibinfo {author} {\bibfnamefont {J.~C.}\ \bibnamefont
  {Dyre}},\ }\bibfield  {title} {\enquote {\bibinfo {title} {A review of
  experiments testing the shoving model},}\ }\href@noop {} {\bibfield
  {journal} {\bibinfo  {journal} {J. Non-Cryst. Solids}\ }\textbf {\bibinfo
  {volume} {407}},\ \bibinfo {pages} {14--22} (\bibinfo {year}
  {2015})}\BibitemShut {NoStop}%
\bibitem [{\citenamefont {Pedersen}\ \emph {et~al.}(2008)\citenamefont
  {Pedersen}, \citenamefont {Bailey}, \citenamefont {Schr{\o}der},\ and\
  \citenamefont {Dyre}}]{ped08}%
  \BibitemOpen
  \bibfield  {author} {\bibinfo {author} {\bibfnamefont {U.~R.}\ \bibnamefont
  {Pedersen}}, \bibinfo {author} {\bibfnamefont {N.~P.}\ \bibnamefont
  {Bailey}}, \bibinfo {author} {\bibfnamefont {T.~B.}\ \bibnamefont
  {Schr{\o}der}}, \ and\ \bibinfo {author} {\bibfnamefont {J.~C.}\ \bibnamefont
  {Dyre}},\ }\bibfield  {title} {\enquote {\bibinfo {title} {{Strong
  Pressure-Energy Correlations in van der Waals Liquids}},}\ }\href@noop {}
  {\bibfield  {journal} {\bibinfo  {journal} {Phys. Rev. Lett.}\ }\textbf
  {\bibinfo {volume} {100}},\ \bibinfo {pages} {015701} (\bibinfo {year}
  {2008})}\BibitemShut {NoStop}%
\end{thebibliography}
\end{document}